\documentclass[letterpaper,twocolumn,10pt]{article}
\usepackage{usenix}
\AtBeginDocument{%
  }

\usepackage{tabularx}
\usepackage{makecell}
\usepackage{multirow} 
\usepackage{comment}
\usepackage{hyperref}
\usepackage{xurl}
\usepackage{graphicx}
\usepackage{booktabs}
\usepackage{xcolor}
\usepackage{tikz}
\usepackage{xspace}
\usepackage{caption}
\usepackage{adjustbox}

\def\authnote{1}

\usepackage{setspace}

\usepackage{amssymb}

 \newcommand{\secref}[1]{Section~\ref{#1}}
\newcommand{\figref}[1]{Figure~\ref{#1}}
\newcommand{\tabref}[1]{Table~\ref{#1}}
\newcommand{\apref}[1]{Appendix~\ref{#1}}

\def\authnote{1}

\newcounter{mynote}[section]
\newcommand{\notecolor}{blue}
\newcommand{\thenote}{\thesection.\arabic{mynote}}
\newcommand{\tnote}[1]{\ifnum\authnote=1\refstepcounter{mynote}{\bf \textcolor{\notecolor}{$\ll$TomR~\thenote: {\sf #1}$\gg$}}\fi}

\newcommand{\checkthis}[1]{\ifnum\authnote=1{\textcolor{cyan}{[CHECK: #1]}}\fi}
\newcommand{\fixme}[1]{\ifnum\authnote=1{\textcolor{red}{[FIXME: #1]}}\fi}
\newcommand{\fixed}[1]{\ifnum\authnote=1{\textcolor{violet}{[FIX: #1]}}\fi}
\newcommand{\fm}[1]{\ifnum\authnote=1{\textcolor{red}{\bf [#1]}}\fi}
\newcommand{\better}[1]{\ifnum\authnote=1{\textcolor{violet}{[BetterWord: #1]}}\fi}
\newcommand{\todo}[1]{\ifnum\authnote=1{\textcolor{red}{[TODO: #1]}}\fi}
\newcommand{\point}[1]{\ifnum\authnote=1{\textcolor{gray}{/* #1  */}}\fi}

\newcommand{\alaa}[1]{\ifnum\authnote=1\refstepcounter{mynote}{\bf \textcolor{teal}{$\ll$Alaa~\thenote: {\sf #1}$\gg$}}\fi}

\newcommand{\eric}[1]{\ifnum\authnote=1\refstepcounter{mynote}{\bf \textcolor{red}{$\ll$EZ~\thenote: {\sf #1}$\gg$}}\fi}

\newcommand{\sarah}[1]{\ifnum\authnote=1\refstepcounter{mynote}{\bf \textcolor{blue}{$\ll$SC~\thenote: {\sf #1}$\gg$}}\fi}

\newcommand{\anote}[1]{\ifnum\authnote=1\refstepcounter{mynote}{\bf \textcolor{purple}{\noindent $\ll${AD~\thenote:} {\sf #1}$\gg$}}\fi}

\newcommand{\elissa}[1]{\ifnum\authnote=1\refstepcounter{mynote}{\bf \textcolor{violet}{$\ll$ER~\thenote: {\sf #1}$\gg$}}\fi}

\newlength{\saveparindent}
\newlength{\saveparskip}
\newcounter{ctr}

\begin{document}

\title{Dual-Use AI Face Swap Apps Are Mostly Unsafe: A Systematic Safety Audit}

\author{
 {\rm Alaa Daffalla}\\
 Cornell University
 \and
 {\rm Sarah Chao}\\
Georgetown University
 \and
 {\rm Eric Zeng}\\
 Georgetown University
}

\maketitle

\begin{abstract}

AI-based image editing tools, such as face swapping algorithms, can be used to transform a clothed image of a person into a sexually explicit image of that person.
These tools are made easily accessible to non-expert users through mobile apps, and have been linked to reports of image-based sexual abuse and cyberbullying
involving synthetic non-consensual intimate imagery.
Apple and Google have begun to remove ``nudification'' apps from their platforms: apps that are marketed with the capability to ``undress'', ``nudify'', or create nude face swaps from images of people. 
However, AI image editing apps that have the same underlying capabilities, but do not present as nudification apps could be also abused to create non-consensual explicit images.
In this paper, we investigate whether AI face swap apps for iOS and Android implement safety measures to prevent the creation of SNCII. 
We identified and downloaded 420 face swap apps, and manually tested 155 eligible apps to see whether they would permit the user to create face swaps with nude images. Our evaluation shows that 70\% of apps with face swap functionality have no technical safeguards against generation of nude images. 
Additionally, we investigated whether face swap apps' descriptions, terms of service, or privacy policies addressed harmful uses of the app, finding that no apps self-describe as nudification apps, but that the majority do not have specific terms of service provisions prohibiting this kind of use. Our findings suggest that to mitigate the threat of UI-bound SNCII threats, platforms and lawmakers must implement policies to mandate safety filters in dual-use AI image editing applications like face swap apps.

\textcolor{red}{Content warning: This paper includes descriptions of applications that can be used to create synthetic nonconsensual
intimate imagery, as well as screenshots of app pages containing potentially explicit content.}

\end{abstract}

\section{Introduction}
Synthetic non-consensual intimate imagery (SNCII) is an increasingly common form of image-based sexual abuse. Perpetrators can take any non-explicit image of a person (e.g., from social media), and use an AI image editing app to create an explicit image depicting that person. These images can be involved in abuse such as cyberbullying, extortion, and intimate partner violence, and can cause victims to experience trauma, anxiety, and stigmatization~\cite{ajay2026psychImplications}.

Recent reports show that SNCII creation and harassment is becoming widespread.
A report from Wired and the Indicator found that since 2023, there have been publicly reported incidents of SNCII being shared in 90 schools, involving over 600 students
\cite{burgess2026deepfake}.
In December 2026, X users discovered that the Grok AI model could be used to undress images of women posted on X, which resulted in an estimated peak of 6,700 SNCII generations per hour before the functionality was paywalled~\cite{danastasio2026grok}.

Prior work and reporting has called attention to ``nudification'' apps as one source of these images: apps whose advertised function is to ``nudify'' existing images, by undressing, swapping faces onto nude bodies, or generating an image of a person in an explicit act~\cite{gibson2025nudification,lakatos2025nudifiers,ttp2026nudify}.
These apps were available on app stores like the Google Play Store, Apple App Store, and the web, and were easily usable by UI-bound users, i.e. did not require expertise in computer vision to use~\cite{gibson2025nudification}. 
While Apple and Google have taken steps to remove nudification apps in response to media attention~\cite{maiberg2024apple}, some still remain available on their platforms~\cite{ttp2026steering}.

But even if Apple and Google remove easily identifiable nudification apps, motivated perpetrators may still be able to find apps that allow them to create SNCII. The underlying capabilities offered by nudification apps are based on common computer vision tools that are also packaged in benign-presenting AI image applications. For example, face swap algorithms used in nudification apps are also available in face swap apps advertised for creating funny photos with friends. ``Undress'' features are based on inpainting techniques, which are branded as ``generative fill'' on Photoshop~\cite{adobe2024inpainting}. Prompting-based attacks like those observed in Jan 2026 on X, where users provided prompted Grok to undress images of women from posts on X, uses image prompting capabilities available on most multimodal foundation models.

While some prominent AI image editing and generation tools, like Gemini, have explicit policies and technical safeguards against creating sexual content~\cite{google2026gemini}, others may or may not include these restrictions. For example, Mink et al. \cite{mink2026unlimited} found that sexual content creators generated visual content using open source models such as Stable Diffusion \cite{podell2024sdxl}, which more recently updated their acceptable use policy to prohibit SNCII creation \cite{Acceptab83:online}.

In this paper, we investigate whether \textit{``dual-use'' face swap apps} are common: apps that offer the face swapping functionality of nudification apps for creating SNCII, and do not have safety policies preventing the creation of explicit images, but are packaged as benign image editing tools. We focus on face swap apps in particular because prior work has found it to be a common method for nudification~\cite{gibson2025nudification,han2025mrdeepfakes}, and because apps with ``face swap'' in their titles and descriptions are common on the iOS App Store and Google Play Store. 
First, we ask:

\begin{itemize}
\item \textbf{RQ1:} Do face swap apps implement technical safety measures to prevent users from creating explicit content?
\end{itemize}

In addition to evaluating whether it is possible to produce explicit outputs with face swap apps, we also investigate whether face swap apps tacitly promote unsafe usage of their apps through how they market themselves on the app store, or their usage or privacy policies. If they do, platforms could use this information as signals for automatically identifying unsafe, ``dual-use'' image editing apps, and app developers could potentially be held liable for enabling harms. We ask:

\begin{itemize}
\item \textbf{RQ2:} Do the app store descriptions of face swap apps indicate whether an app can create SNCII?
\item \textbf{RQ3:} Do the terms of service of face swap apps prohibit users from creating SNCII?
\item \textbf{RQ4:} Do the privacy policies and practices of face swap apps benefit users who want to hide SNCII creation?

\end{itemize}

To answer these questions, we conducted a systematic safety evaluation of face swap apps on the iOS App Store and Google Play Store. We identified 420 face swap apps on the iOS App Store and Google Play Store, and evaluated 155 eligible apps to see whether they implemented technical safety measures to prevent the creation of SNCII. Using four AI-generated image pairs, we manually tested whether the app permitted users to swap a face onto a nude body (see Appendix~\ref{sec:ethics} for a detailed discussion of the ethical considerations of this method). Additionally, we performed a mixed-methods analysis of the app descriptions, terms of service, and privacy policies, to characterize apps' policies regarding SNCII. Among our contributions:

\begin{itemize}
\item We show that 80\% of face swap apps on the Apple App Store and 59\% of apps on the Google Play Store allow nude face swaps without restrictions (Section~\ref{sec:results_swap}).

\item We find that 68.3\% of face swap apps are
bundled with other features for creating deepfakes from images, such as image-to-video and voice synthesis, which could be used to create multimodal SNCII (Section~\ref{sec:app-descriptions}). 

\item We demonstrate a method for manually verifying safety guardrails for image editing using AI-generated nude images, along with ethical implications and alternatives that we considered (Appendix~\ref{sec:ethics}). We offer suggestions to future researchers and trust \& safety operators for using this process as part of AI safety evaluations.

\item We analyze the space of legal, policy, and technical options to address unsafe, ``dual-use'' face swap apps, finding that legal avenues for restricting software for creating SNCII are limited in the U.S., putting the onus on platforms like Apple and Google to address this problem (Section~\ref{sec:discussion_law}). 
\end{itemize}

\section{Background and Related Work}

\paragraph{Image-based sexual abuse.} Image-based sexual abuse (IBSA) is a broad term that refers to the creation and distribution of nude or sexual imagery without the subject's consent \cite{mcglynn2017image, powell2018image}. 
IBSA encompasses a wide range of harms \cite{mcglynn2017beyond}, including but not limited to: sextortion \cite{cross2025pay, henry2015beyond, roberta2022cyber}, non-consensual pornography (NCP) or also known as ``revenge porn'' \cite{eaton20172017, kirchengast2019legal, konstantinos_papachristou_revenge_2024}, non-consensual intimate imagery (NCII), and synthetic NCII (SNCII) among others. In SNCII, perpetrators leverage AI tools and technologies to create explicit or sexualized depictions of a real person (i.e., ``deepfakes'').

\paragraph{Deepfakes and SNCII.}
Over the last decade the prevalence of deepfakes --- media manipulated or created with AI technologies --- has increased  significantly, particularly for image-based abuse \cite{alena2025deepfakes}. A 2024 report has concluded that 96\% of deepfakes are non-consensual sexual or intimate deepfakes (SNCII) \cite{baekgaard2024technology}. We adopt a broad definition of deepfake as synthetic media that manipulates a real person's likeness to construct a false but believable narrative \cite{chesney2019deep, vaccari2020deepfakes, TOLOSANA2020131}.

The underlying computer vision methods enabling SNCII have also grown in variety and accessibility. Early deepfake generation relied on face swapping Generative Adversarial Network (GANs) \cite{ganin2016deepwarp, kim2018deep, radford2015unsupervised} to replace one person's likeness onto another. More recent approaches include inpainting \cite{pathak2016context, yeh2017semantic, li2017generative}; replacing regions of an image with AI-generated content, as well as Low-Rank Adaptation (LoRAs) \cite{hu2022lora} fine-tuning; adapting general purpose diffusion models to a specific person's likeness with minimal training data.  

Generative AI tools leveraging the above methods have been used extensively to create synthetic intimate imagery \cite{mccosker2024making, umbach2024non, winter2020deepfakes, wei2025we}, and AI-generated child sexual abuse material (CSAM) \cite{o2026ai, thiel2023generative, AIbecomi99:online}. More recently, Grok was brought under the spotlight for enabling the creation of SNCII depicting women and minors \cite{Hundreds70:online}. 
Distribution channels and online communities have also been studied in an effort to deplatform and limit the abuse \cite{timmerman2023studying, han2025characterizing}.

Several works have illuminated broader patterns in SNCII and IBSA perpetration and victim perception \cite{fido2022celebrity, her2025mitigating, ruvalcaba2020nonconsensual, bloom2014no}. 
For example, it has been reported that 94\% of deepfake porn targets are people in the entertainment industry or ``celebrities'' \cite{2023Stat68:online}. Some works \cite{fido2022celebrity, her2025mitigating} found that the celebrity status of SNCII victims elicits more lenient public judgments of harm and criminality, suggesting that public visibility normalizes image-based sexual abuse deterring victims from help seeking. In a similar vein, recent work by Henry et al. \cite{henry2026imageabuse} shows that IBSA perpetration is embedded in gendered norms and power structures that normalizes and exacerbates abuse.

\paragraph{Investigating SNCII in the wild.} Early SNCII creation required significant technical expertise. Timmerman et al. \cite{timmerman2023studying} found how the originating deepfake reddit community on $r/deepfakes$ and its successor platform, the MrDeepFakes relied on Github hosted tools such as DeepFaceLab \cite{iperovDe33:online} requiring users to source training data, process facesets, and train models. However, these barriers have since collapsed, increasingly making it easier for UI-bound perpetrators --- interact with systems via standard UIs to cause harm --- to create SNCII. For example, Gibson et al. \cite{gibson2025nudification} conducted a systematic study of 20 AI nudification websites and found that websites enable UI-bound SNCII with easy to locate nudification features surfaced through a regular app walkthrough. Similarly, the Tech Transparency Project has surfaced 31 apps across both the Apple and Google app stores that can be used to create SNCII \cite{ttp2026steering}.  

More recently, Brigham et al. \cite{brigham2026examining} investigated the security and safety risks of AI companion apps --- conversational apps for chatbot-human interaction. They found that these apps allow users to upload real photos of any person to generate a companion without the depicted person(s)' consent. This along with other built-in sexual interaction features provides a direct pathway to SNCII. Their work further shows that SNCII is no longer limited to deepfake platforms and nudification apps but is embedded in mainstream, app store accessible AI products further lowering the barrier for SNCII. 

\paragraph{Investigating SNCII for face swapping apps.}
While prior work has examined dedicated deepfake communities \cite{timmerman2023studying}, purpose-built nudification websites \cite{gibson2025nudification}, and AI companion apps \cite{brigham2026examining}, the face swapping app ecosystem has received comparatively little safety scrutiny. A recent investigation by the Tech Transparency Project \cite{ttp2026nudify} found that multiple apps, including RemakeFace, a face swapping app marketed for entertainment with over 5 million downloads on Google Play can be used to generate nudified images of real people, illustrating the dual-use risk embedded in app store products. We build on this by conducting a systematic, large scale safety evaluation of face swapping mobile apps, examining whether their features and safeguards prevent misuse for SNCII creation.

\section{Methods}
\label{sec:methods}

Next, we describe our methodology for evaluating the safety of face swap apps and analyzing their policy documents. In summary, we 1) compiled a list of relevant face swap apps from the Apple App Store and Google Play Store using the search function, 2) scraped their descriptions, privacy policies, and terms of service, 3) downloaded each app and attempted to conduct face swaps using AI-generated nude test images, and 4) qualitatively analyzed the policy documents and descriptions. 

\subsection{App Data Collection}

\paragraph{Keyword search for face swap apps.} We obtained a list of apps from both the Apple App Store and the Google Play Store using keyword search. Specifically, we searched for the term \textit{``face swap''} and collected all results that showed up during the search until reaching end of page. We did not include any sponsored or advertised results. In total, we collected around 210 apps from the App Store and 250 apps from the Play Store. 

\paragraph{Collecting app store metadata.}
For each app, we scraped metadata from their respective app stores: app descriptions, developer name, ID, URL, and website, app categories/genres, OS and device compatibility, app ratings and installs, 
app scores, versions, content ratings, header images, and supported languages. We did not analyze app reviews because our goal was to identify signals that could predict or inform decisions by platforms when reviewing app submissions prior to deployment.    
We used a Node.JS open source library for scraping app metadata for both the App Store \footnote{https://github.com/facundoolano/app-store-scraper} and Play Store \footnote{https://github.com/facundoolano/google-play-scraper}.

\paragraph{Collecting privacy policies and terms of service.}
After collecting data from the app store listings, we scraped the privacy policies and terms of service of each app, if available. First, we obtained the URLs. For privacy policies, we were able to extract the URLs from the app store metadata, as both stores require that app developers provide a URL on their store page. On the other hand, terms of service are not a required field. To find those URLs, we searched the app's descriptions for links to terms of service or related documents, using Gemini 2.5 Flash. For apps that did not provide a terms of service URL in the description, we manually searched for terms of service documents on app developers' websites.

Once we obtained URLs for the documents, we scraped the content at the URLs, and saved them to a text file using a web crawler based on Puppeteer. In cases where Puppeteer was unable to scrape the documents (e.g. for PDFs or Google Docs), we manually opened and copied the documents into a text file. In total, we collected 353 privacy policies (167 iOS, 186 Android) and 250 terms of service (169 iOS, 81 Android). Reasons for missing privacy policies and terms of service include apps that did not include a URL, inaccessible URLs, and documents in other languages.

\subsection{Face Swap Safety Evaluation}

\begin{figure*}[t]
    \centering
    \includegraphics[width=0.9\textwidth]{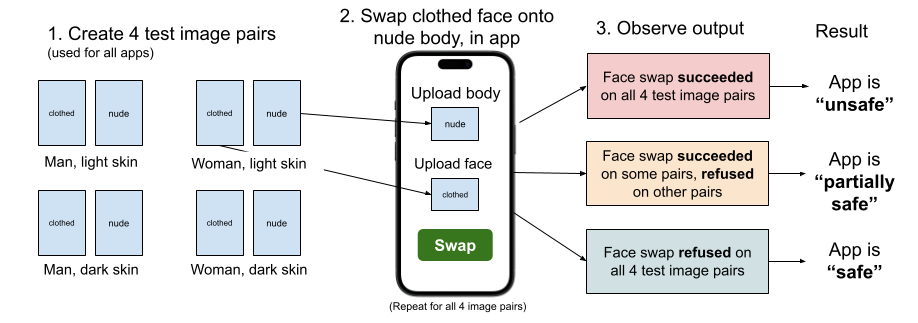}
    \caption{An overview of our safety evaluation procedure.}
    \label{fig:test_method_diagram}
\end{figure*}

To test whether face swap apps permit the creation of SNCII, we developed a structured testing methodology: generating a set of test images, preparing a test environment, manually testing each app, and recording the results of the testing procedure. \figref{fig:test_method_diagram} shows an overview of our procedure.

\subsubsection{Threat model.} We begin with an explanation of the threat model which guided our testing procedure. We model a UI-bound attacker \cite{freed2018stalker} that interacts with face swapping applications exclusively through standard user interfaces. This is consistent with the increasingly low technical barrier to SNCII creation documented in prior work \cite{gibson2025nudification, brigham2026examining, ttp2026nudify}. The attacker's goal is to generate SNCII of a victim through the app's available features, by swapping a victim's face image into a nude body image, with both images supplied by the attacker. Importantly, the attacker's objective is not to adversarially stress test the service's safety mechanisms, but to evaluate whether those mechanisms are present and effective in preventing SNCII. 

\subsubsection{Safety testing.}

Next, we describe our procedure for determining if face swap apps included safety measures to prevent face swaps with explicit images.  

\paragraph{Test images for face swaps.}
\label{sec:test_images}
A single test requires a pair of images: a (clothed) target face, and a (nude) target body. The target face is to be swapped onto the target body. We determined that we should test 4 pairs of images to account for variations that might affect safety measures like nudity detection \cite{tariq2019review, lopes2009bag}. Prior work has indicated that nudity detection performance is affected by differences in skin tone and sex \cite{riccio2024art, trinh2021examination, akyon2023state}. Thus, we generated both target face and body images for 
two conditions; a male and female sex, and skin tone; a light and dark skin tone. 

We generated the clothed, target face images using the FLUX.2 image model \footnote{\url{https://bfl.ai/models/flux-2}}, using prompts to specify gender and skin tone (prompts in Appendix~\ref{appendix:test_image_prompts}). For the nude, target body images, 
we retrieved images from an AI pornography website that hosts generated images. We searched for four images, one for each gender and skin tone condition. Each image was photorealistic, and depicted one person, full-body, in a standing position, and nude. To protect the researchers well-being during the course of the testing, we avoided images that depicted explicit sexual acts.


\paragraph{Test environment.}

Our testing environment included multiple devices and test accounts. All devices and accounts used in this study are synthetic and used for the purposes of this study only. 
We initially used a MacBook Pro laptop (MacOS Sequoia v15.5) and an Android Studio emulator running on a MacBook Pro device (MacOS Sequoia v15.5) respectively. However, we found that some apps were incompatible with these environments, so for those apps we used an iPhone 11 (iOS 26.1) or a Motorola Moto G Play (Android v16). 
We created new Apple IDs and Google Accounts (one for each researcher involved) to access the app stores. Additionally, we used Privacy.com to create virtual payment methods to pay for apps that required subscriptions to use.

\paragraph{Testing procedure.} First, the researcher installed the app and navigated through the UI to find the face swap feature. If necessary to access face swap functionality, the researcher would pay for a subscription or credits. Once the face swap feature was located, the researcher uploaded a target body image and target face image from an image pair, and swapped faces. This step is repeated four times.

For each app we recorded whether the face swap was successful; resulted in an output image of the target face on the target body or not. We also captured refusals where the face swap would fail, and other technical errors, such as when apps crashed. If the SNCII swap resulted in an ambiguous generic error that did not indicate that the uploaded content violated safety rules, we attempted a face swap with only clothed images. If the face swap with clothed images succeeds then we marked the SNCII swap as safe (because it blocked nude inputs). Otherwise, we marked the app as out of scope due to a technical failure. 
If the app did not allow us to swap faces with two arbitrary images, we checked whether the app offered a template-based swap feature,  i.e. app developer-provided target body images, and whether they were sexually suggestive or explicit. 

\subsubsection{Apps Excluded from Safety Tests}

265 of the 420 apps in our initial dataset were not able to be tested according to our safety testing procedure. We describe the reasons why, and show an overview of the included and excluded apps in Table~\ref{tab:out-scope}.

\paragraph{Lacks two-image face swap feature.}
In our initial safety tests, we found that not all app had functionality that were compatible with our testing procedure. Instead of swapping the face from one image to the body of a second image, apps had other modalities: template-based face swaps (users can only upload a target face, the target body is a developer-provided image such as a movie scene, professional headshot -- see \figref{fig:description-templates} in \apref{appendix:codebooks}), video and live swaps (swapping faces into a video or live capture), prompt-based AI image or video generation, and face retouching features.

However, our primary research question is to investigate whether face swap apps can be used to generate SNCII targeted at a particular individual. So, we established an inclusion criteria that determines which apps are in scope for safety evaluation: the app must allow for targeted identity face swaps, meaning the app must take as input two images, a target face image, and target body image that are both arbitrary and provided by the user. 
Furthermore, we restricted our criteria to image-based face swaps, as our test data were images.

165 apps were excluded for not supporting face swaps between two arbitrary images. 
50 apps (App=28, Play=22) were excluded because they did not allow a custom target body to be uploaded, only supporting swaps with target bodies that the developer provided.  
Another 21 apps were excluded (App=11, Play=10) because they only provided a video face swap feature, which we excluded because we did not create video versions of our test image pairs.
An additional 10 (App=6, Play=4) because they only allowed faces to be swapped onto live video captured on camera.
Two iOS apps were excluded because they only support group picture face swaps: these apps take a single image containing multiple people as input, and swap faces between the people in the single image. 
82 apps were excluded because they did not contain a face swap feature.

\paragraph{App technical errors.}
Additionally, we excluded apps that were not available or not functional on our test device.
For 43 apps, the app store listings were no longer available at the URL we originally scraped. This means that the app might have been taken down or moved to different URL between our metadata scrape and our test. Additionally, for 38 apps, we encountered technical failures while attempting to use the app. For example, many apps would not launch, crash, or show continuous ads that prevented access to functionality. Lastly, 19 apps were excluded because they were not in English, or were outdated and incompatible with the test device. 

In the end, we conducted our safety evaluations on \textit{155 apps}.
We refer to these 155 apps for the remainder of the paper.

\begin{table}[h]
\centering

\begin{tabular}{lcccc}
\toprule
 & \textbf{iOS} & \textbf{Android} & \textit{Total}\\
\midrule
\multicolumn{3}{l}{\textit{Scope}} \\
\quad Apps          & 210 & 210 & 420 \\
\quad In scope       & 85 & 70  & 155\\
\quad Out of scope   & 125 & 140 & 265\\
\midrule
\multicolumn{3}{l}{\textit{Reason for exclusion}} \\
\quad URL not found      & 16 & 27 & 43 \\
\quad Template-based     & 28 & 22 & 50\\
\quad Technical failure  & 19 & 19 & 38\\
\quad Live swap          & 6 & 4 & 10\\
\quad Video swap         & 11 & 10 & 21\\
\quad Group swap         & 2 & 0 & 2\\
\quad No face swap       & 39 & 43 & 82\\
\quad Other              & 4 & 15 & 19\\
\bottomrule

\end{tabular}

\caption{Summary of apps included and excluded from our safety testing procedure}

\label{tab:out-scope}
\end{table}

\subsection{App Metadata Analysis}
In this section we discuss how we analyzed the apps' metadata: app store descriptions, terms of service, and privacy policies.  

\begin{figure*}[t]
     \centering
     \includegraphics[scale=0.45]{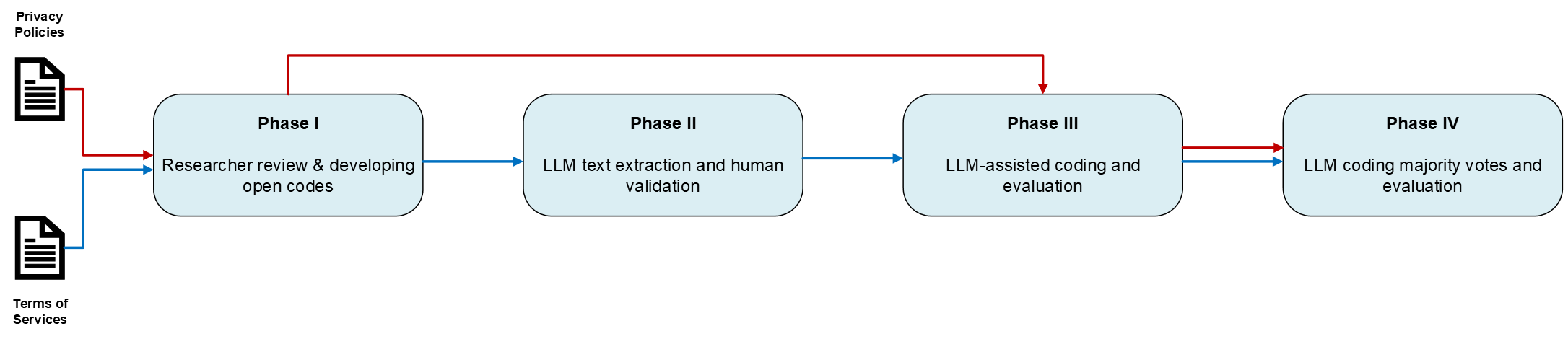}
     \caption{The four phases for analyzing the privacy policies and terms of service documents for all apps. Unlike for terms of services, we did not leverage LLM extraction for privacy policies because we created \textit{apriori} codes.}
     \label{fig:tosanalysis}
\end{figure*}

\subsubsection{Descriptions}
To understand how app developers advertise face swap features and what motivation they provide for its use, we analyzed the apps' descriptions from their app store listings. 
Two researchers reviewed and open coded the descriptions of 20 iOS apps. In this initial round, the researchers identified multiple areas of interest, including what use cases apps promote for face swapping, and what other features were offered beyond face swaps. Following this, the researchers met to resolve differences in the definitions and applications of codes in the overlapping set of 20, resulting in 16 open codes across two categories (see Appendix~\ref{appendix:codebooks}). One researcher then coded the remainder of the app descriptions.

\subsubsection{Terms of Service}
\label{sec:tos_methods}
We performed an analysis of the terms of service documents across all apps in four phases (see \figref{fig:tosanalysis}). The goal of this analysis was to identify to what extent app developers discuss or prohibit creation of SNCII through their apps. We took a bottom-up approach to the analysis that involved researcher review, developing open codes and using LLM-assisted document extraction and qualitative coding. 

In the first phase, two researchers read through a sample of 10 terms of service documents and created initial codebooks. The codebook centered on two themes: prohibited uses that could be construed as SNCII, and user consent for being depicted in images. They met to merge codes and this phase resulted in a codebook with eight open codes (see \tabref{tab:tos-codebook} in Appendix~\ref{appendix:codebooks}). 

Because terms of service documents are long and contain extraneous information, in the second phase we used an LLM (Gemini 2.5 Flash) to extract the subsections of text relevant to our codebook. We used two different prompts (\tabref{tab:prompts-extraction}) to extract content pertaining to prohibited content and user consent. We ran our script on all terms of service documents to extract the relevant sections. To manually validate the output, two researchers each manually extracted sections of prohibited content and user consent across 10 terms of service documents. While the validation confirmed that the LLM was not hallucinating or adding irrelevant content we found that on the other hand the LLM was omitting user consent sections and we attributed this to how the prompt was structured. We iteratively refined the prompt to better capture the notion of requiring consent for use of images for depicted persons until we reached a recall of 91\%.

In the third phase we qualitatively coded the extracted sections. An initial batch was coded by the researchers, and the remainder was automatically coded using Gemini 2.5 Flash. 
First, two researchers independently coded the extracted sections of 10 terms of service documents. 
The researchers resolved disagreements, updating the codebook if definitions needed clarification. Then, the codebook was converted into prompts, and the LLM was used to code the same set of 10 documents. Agreement was calculated between the human coders and the LLM using Fleiss' Kappa. We reviewed the results and iterated on the prompts until reaching good agreement with the LLM ($\kappa=0.91$). To further validate the LLM coding performance, one researcher coded a new set of 10 documents and computed iterrater reliability with the LLM codes, achieving an agreement level of $\kappa=0.77$.  

To further improve the LLM coding performance, in the final phase we averaged the LLM codes over multiple runs. Multiple works in the literature have suggested this as a best practice for improving LLM performance on reasoning tasks \cite{wang2022self}. As such, for each extracted terms of service document, the LLM would code the document three times and then we calculated a majority vote for each code. We then calculated the inter-coder agreement between the LLM majority vote and the researchers' code on the second set of 10 documents resulting in a $\kappa=0.829$.

\subsubsection{Privacy Policy}
\label{sec:privacy_methods}
We qualitatively analyzed privacy policies to identify privacy protections that face swap app developers offer to its users, potentially to shield the developer or user from legal consequences of creating SNCII face swaps.
Two researchers read through the privacy policies, and created an initial codebook of \textit{a priori} codes based on privacy policy literature. The codes cover: is face data collected, where is it processed, how long is it retained, and do the developers use it for other reasons?

Following this, two researchers independently coded 15 policies, and iteratively resolved disagreements and updated the codebook. Then, one researcher implemented the codebook as prompts for the LLM (\tabref{tab:pp-codebook}), and iteratively refined the prompts by comparing the outputs to the researchers' codes, and adjusting the prompts until it matched the researchers' codes. 
Then, to validate the LLM output, the researcher coded an additional 15 policies, and computed interrater reliability with the LLM's output, achieving an agreement of 0.90 with Krippendorff's Alpha.

\subsection{Ethics}
We provide a detailed ethical considerations section in Appendix~\ref{sec:ethics}, but given the ethically fraught nature of using AI-generated nude images, we briefly discuss our ethical considerations. In order to evaluate the safety of face swapping apps we had to generate a set of nude images for the testing procedure. After considering several alternatives, and the anticipated harms for each, we concluded that AI-generated nude images that we verified did not depict a real person would pose the least risk to individuals. We also discuss responsible disclosure and offsetting our in-app purchases with a donation to a non-profit organization.

\section{Results}

In this section we discuss our safety testing results and present an analysis of apps' descriptions, terms of services, and privacy policies.

\subsection{Face Swap Safety Evaluation (RQ1)}
\label{sec:results_swap}
First, we present results on whether face swap apps permit users to swap faces onto nude bodies targeting a particular individual (see \figref{fig:test_method_diagram}). We find that the majority of apps across both the App and Play stores result in successful SNCII face swaps indicating that most of these apps are \textit{unsafe} and do not have nudity filters.

\begin{table}[h]
\centering

\label{tab:platform}
\begin{adjustbox}{max width=\columnwidth}
\begin{tabular}{lccc}
\toprule
 & \textbf{iOS} & \textbf{Android} & Total\\
\midrule
\multicolumn{3}{l}{\textit{Safety (in-scope only)}} \\
\quad Unsafe  & 68/85 (80.0\%) & 41/70 (58.6\%)  & \textbf{109}\\
\quad Partially safe  & 7/85 (8.24\%) & 9/70 (12.86\%) & \textbf{16}\\
\quad Safe  & 10/85 (11.76\%) & 20/70 (28.57\%) & \textbf{30}\\
\bottomrule

\end{tabular}
\end{adjustbox}
\caption{Safety evaluation of in-scope face swap apps. We show a per platform breakdown for safety outcomes. }

\label{tab:safety}
\end{table}

\begin{table}[h]
\centering
\begin{tabular}{ccccr}
\toprule
\makecell{\textbf{light} \\ \textbf{woman}} & \makecell{\textbf{dark} \\ \textbf{woman}} & \makecell{\textbf{light} \\ \textbf{man}} & \makecell{\textbf{dark} \\ \textbf{man}} & \textbf{Count} \\
\midrule
Safe & Safe & Safe & Unsafe  & 5 \\
Safe & Unsafe  & Safe & Safe & 3 \\
Safe & Safe & Unsafe  & Safe & 2 \\
Unsafe  & Unsafe  & Unsafe  & Safe & 2 \\
Safe & Safe & Unsafe  & Unsafe  & 1 \\
Safe & Unsafe  & Unsafe  & Unsafe  & 1 \\
Unsafe  & Safe & Safe & Safe & 1 \\
Unsafe  & Unsafe  & Safe & Unsafe  & 1 \\

\bottomrule
\end{tabular}
\caption{Safety results across the four image pairs for the 16 partially safe apps. We show the different failure modes for these apps.}
\label{tab:test-pairs}
\end{table}

\subsubsection{Number of apps that allowed explicit face swaps} We found that 109 out of 155 apps are \textit{unsafe} (\tabref{tab:safety}): they permitted face swaps from an image of a clothed person to an image of a nude person, across all four image test pairs we tested (covering the combinations of dark/light skin tone and men/women genders). 
This is a 70\% success rate for explicit face swaps, indicating that the majority of apps across iOS and Android do not prohibit the targeting of particular individuals and creation of nude images that might have been obtained non-consensually.

Sixteen apps were \textit{partially safe} (\tabref{tab:safety}), meaning that face swaps succeeded for at least one, but not all four of the image test pairs we tested. This indicates that the app has some kind of safety mechanism like a nudity filter, but it did not work consistently across images. In \tabref{tab:test-pairs} we show the different combinations of failure modes for the partially safe apps. We find that for 50\% of the failure modes, the failure only occurred in the images of people with dark skin tones. This echoes similar findings in skin tone bias in deepfake detection \cite{trinh2021examination, hazirbas2021casual}.

\begin{figure}[h]
     \centering
     \includegraphics[width=0.45\linewidth]{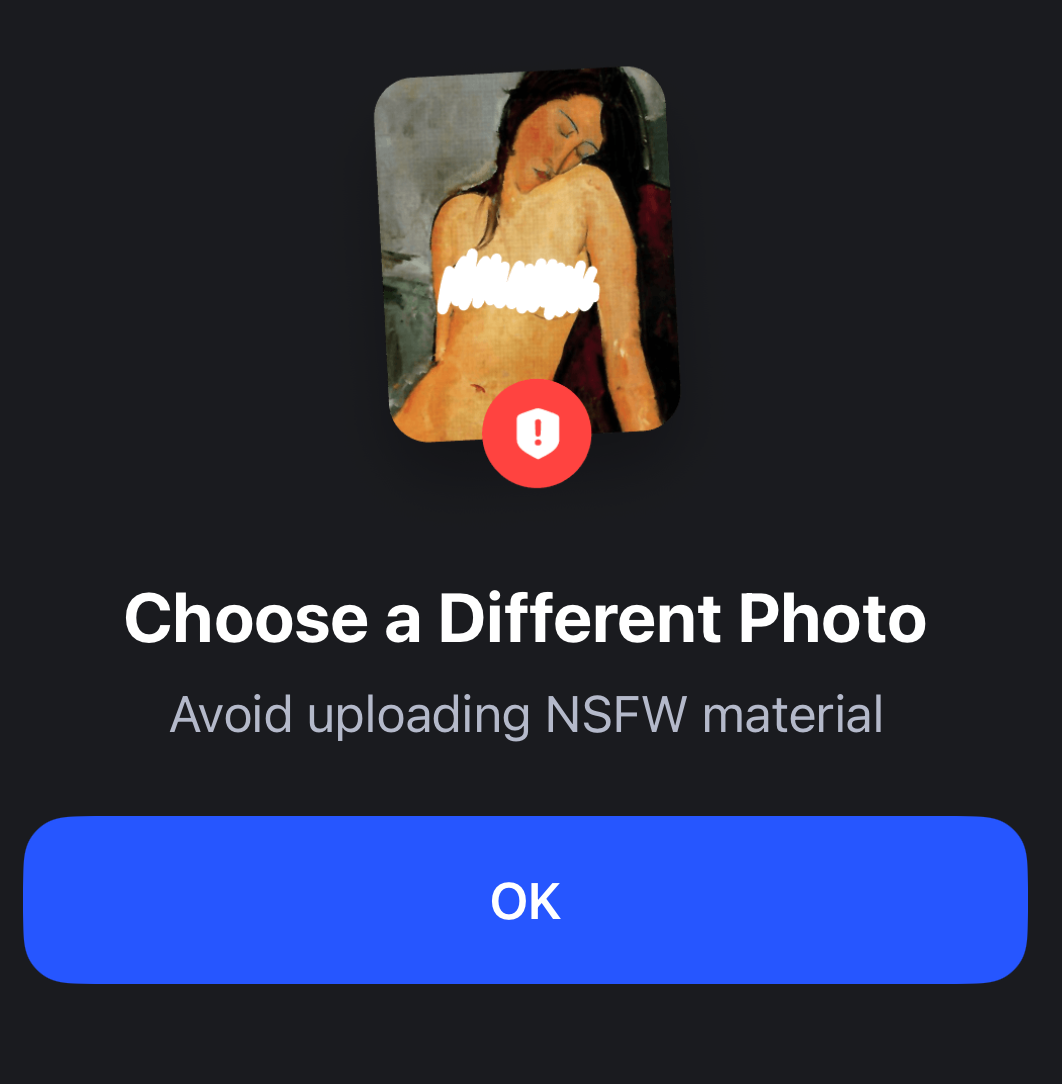}
     \quad
      \includegraphics[width=0.48\linewidth]{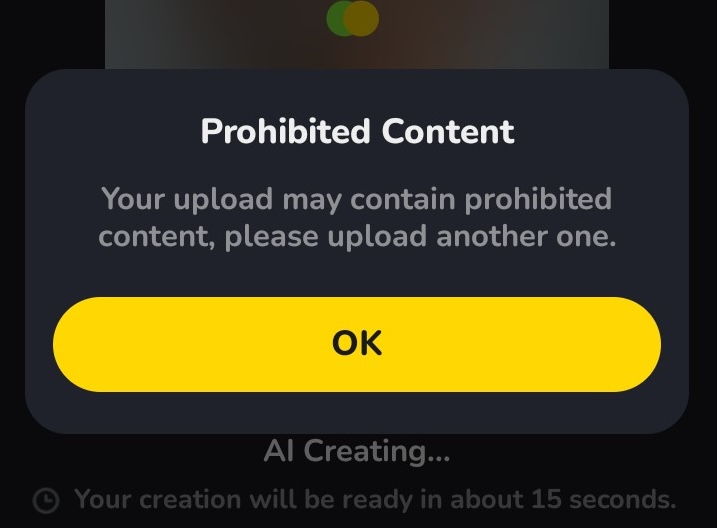}
     \caption{Examples of refusal messages for safe apps.}
     \label{fig:refusals}
\end{figure}

Finally, 30 apps were \textit{safe}: they refused to swap faces on all of the four image pairs we tested. When we attempted to perform the face swap, these apps responded with refusal messages such as flagging the input nude image as being illegal, inappropriate, or not safe for work (NSFW). We show examples in \figref{fig:refusals}.

\subsubsection{Platform differences}
Face swap apps on iOS were more likely to permit face swaps with nudity than face swap apps on Android: 80\% of iOS apps allowed nude face swaps compared to 58.6\% of Android apps (\tabref{tab:safety}). A two-sample Pearson chi-squared test indicated a statistically significant association between platform and safety test result ($\chi^2(2, N=155)=8.9032, p=0.012$). One possible explanation for this difference is policy: the Google Play Store requires apps that provide access to generative AI block generation of restricted content~\cite{google2026aigencontent}, while the App Store does not have a specific policy on safety for apps allow users to use generative AI~\cite{apple2026appreview}.

\subsubsection{App paywalls \& safety}
We find that 46.4\% of in-scope apps were behind a paywall and 
required in-app purchases to access face swapping features (see \tabref{tab:subscription-safety}). In particular, around 50\% of both unsafe and partially safe apps required in-app purchases compared to only 26.7\% safe apps. Overall, the average price to access face swaps was \$5.47 for unsafe apps, \$7.61 for partially safe apps, and \$5.30 for safe apps. 

\begin{table}[h]
\centering

\begin{tabular}{lccc}
\toprule
Safety & Total & Purchase Req'd & Avg.\ Price (\$) \\
\midrule
Unsafe      & 109 & 61 (56.0\%) & 5.47 \\
Partial   & 16  & 8 (50.0\%) & 7.61 \\
Safe    & 30  & 9 (30.0\%) & 5.30 \\

\bottomrule

\end{tabular}

\caption{In-app purchase requirements by app safety outcomes}
\label{tab:subscription-safety}

\end{table}

\subsubsection{Developer country}
The developers of face swap apps are widely distributed across the globe. 
We extracted the country where developers of Android apps are based, using the legal address publicly shown in the Play Store (see Table~\ref{tab:country}). The top countries where developers of Face Swap apps are based are India (18\% of apps), China (14\%), and Vietnam, Hong Kong, and Singapore (7\% each). Only 3 apps were based in the U.S. The wide geographic distribution shows how deployment of AI tools, safe or not, is global, and points to challenges in governance as regulations at the national level may not affect all developers, which we discuss later in \secref{sec:discussion_law}.
\begin{table}
    
\begin{adjustbox}{max width=\columnwidth}
\begin{tabular}{lrrrrr}
\toprule
 &      &        & Partially  &  &  \\
 \textit{Country} & Safe & Unsafe & Safe      & Total & \% Safe \\
\midrule
India & 2 & 9 & 2 & 13 & 15.4 \\
China & 3 & 5 & 2 & 10 & 30.0 \\
Vietnam & 1 & 3 & 1 & 5 & 20.0 \\
Hong Kong & 1 & 3 & 1 & 5 & 20.0 \\
Singapore & 0 & 5 & 0 & 5 & 0.0 \\
Pakistan & 1 & 3 & 0 & 4 & 25.0 \\
Indonesia & 1 & 1 & 2 & 4 & 25.0 \\
United States & 1 & 1 & 1 & 3 & 33.3 \\
Israel & 1 & 1 & 0 & 2 & 50.0 \\
Italy & 1 & 1 & 0 & 2 & 50.0 \\
\bottomrule
\end{tabular}
\end{adjustbox}

\caption{Top 10 countries that developers of Android face swap apps are based in. The iOS App Store does not display the developer's country.}
\label{tab:country}

\end{table}

\subsubsection{Popularity of apps \& safety}
We find that Reface is both the most popular iOS and Android app in our dataset
with the iOS app having 491,584 ratings and the Android version having 
over 100 million installs. Both the iOS and Android versions of Reface were partially safe, failing on the test images depicting the man with darker skin tone. Overall, the mean number of ratings of iOS apps is 9,384 ratings, while that for Android is two million installs. 

We hypothesized that the popularity of face swap apps may correlate with whether they have a safety filter; either that unsafe apps could be more popular, or that popular apps are more likely to be safe because they are more likely to be scrutinized by the app store platform. However, we find that the popularity of an app does not correlate with safety filters. 

\begin{figure}[ht]
     \centering
     \includegraphics[scale=0.7]{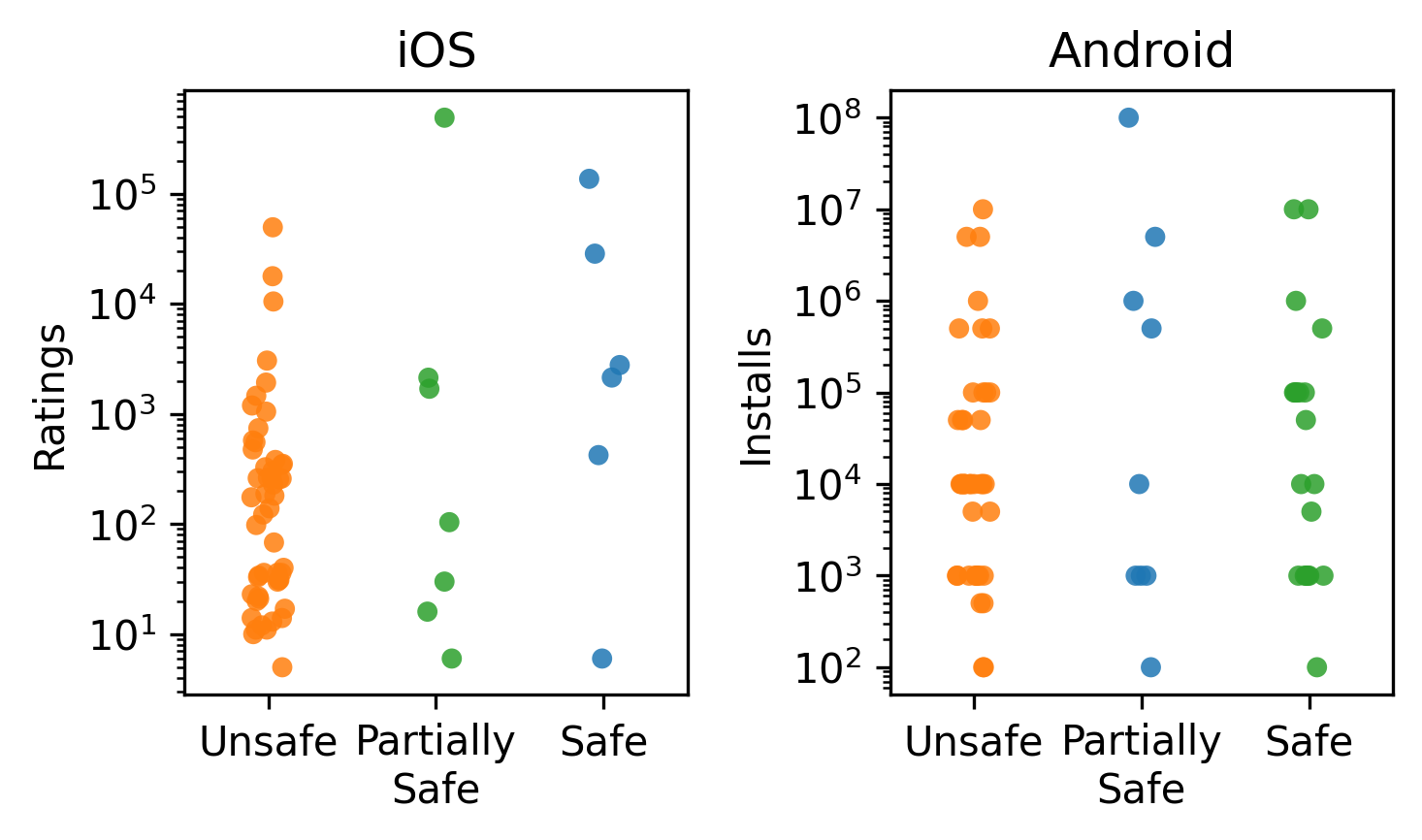}
     \caption{Comparison of app popularity across safety test outcome. There was no significant effect of the number of ratings (iOS) or installs (Android) on popularity. }
     \label{fig:ratings-plot}
\end{figure}

To test this hypothesis, we used the number of app ratings on the App Store and number of installs on the Play Store as proxies for popularity. Apple does not release the number of installs on the App Store, so ratings were used as a measure of usage instead. On the Play Store, the number of installs are grouped by tier (e.g., 1000+, 100,000+). \figref{fig:ratings-plot} shows the distribution of popularity between the test outcomes per platform. Analyses of variance
based on multinomial logistic regression indicated that there was no significant effect of the number of ratings on safety test outcome for iOS ($\chi^2(2,N=65)=5.67, p=n.s.$), nor installs on safety test outcome for Android ($\chi^2(2,N=70)=0.77, p=n.s.$).

\subsubsection{Non-refusal safety measures}
Some apps included weaker safety mechanisms than refusing to perform face swaps. We observed some instances of deterrence messages: warnings to users to not upload or generate NSFW or explicit material. However, in each of these cases, we still successfully completed a face swap, meaning that the warnings were not backed with any actual safety protections. 
One iOS app, Face Swap: AI Photo Generator, produced an output image where the private parts were blurred. We counted this as successful face swap as the images still showed a mostly-nude body.

\subsubsection{Apps with suggestive templates} 
Fifty apps did not meet our inclusion criteria for the above tests  because they offered ``template-based'' face swaps -- i.e., you can only upload a face image and swap it into existing templates (\tabref{tab:out-scope}). During our testing, we qualitatively analyzed whether any of these templates were nude or sexually suggestive, as these could also be used to create SNCII, albeit with less control over the output.
We found that no apps had nude templates, but 22 had sexually suggestive templates. They included descriptions such as \textit{hot girls}, \textit{bikini}, \textit{sexy}, and \textit{AI kiss/hug}. 
Other non-suggestive template categories included beach, Christmas, and birthday.

\begin{figure}[h]
     \centering
     \includegraphics[width=0.35\linewidth]{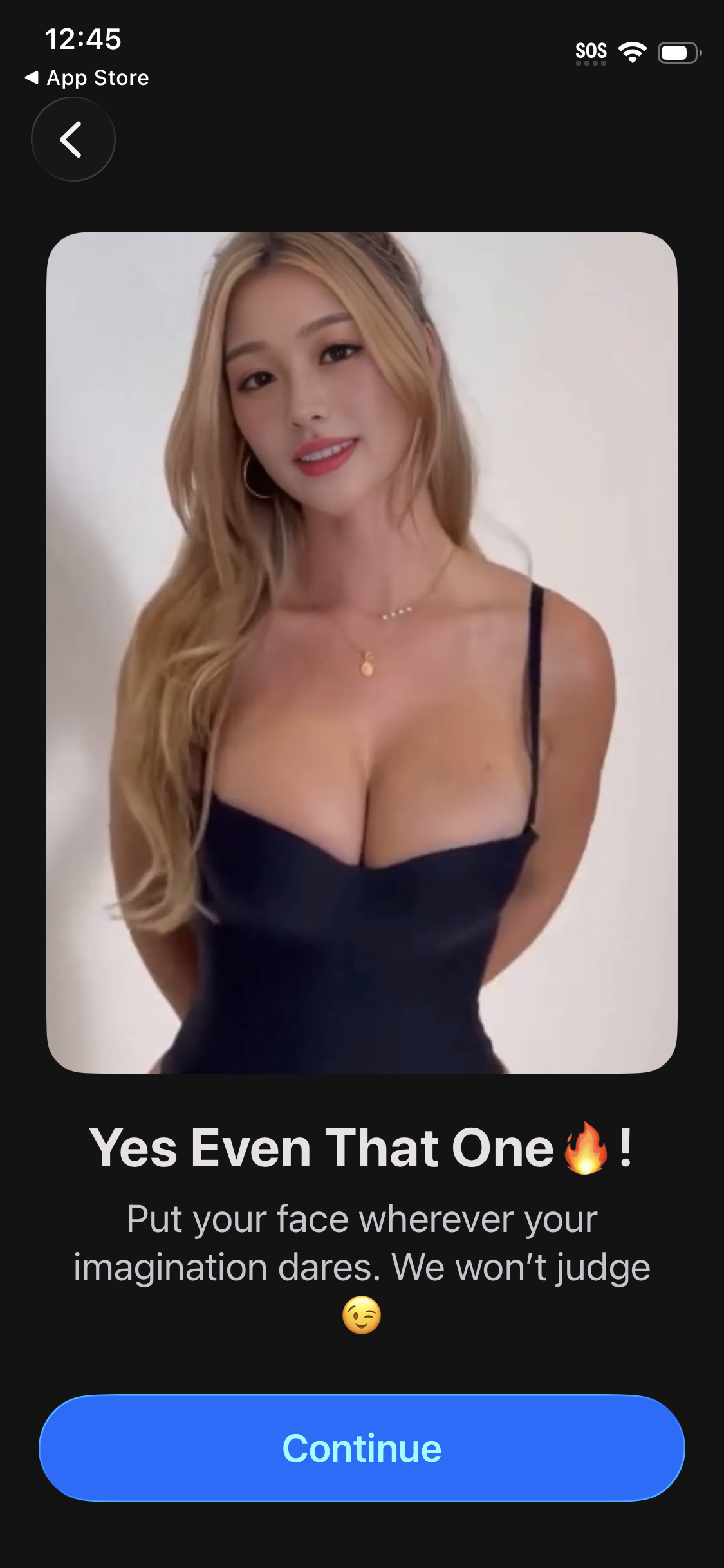}
     \quad
     \includegraphics[width=0.35\linewidth]{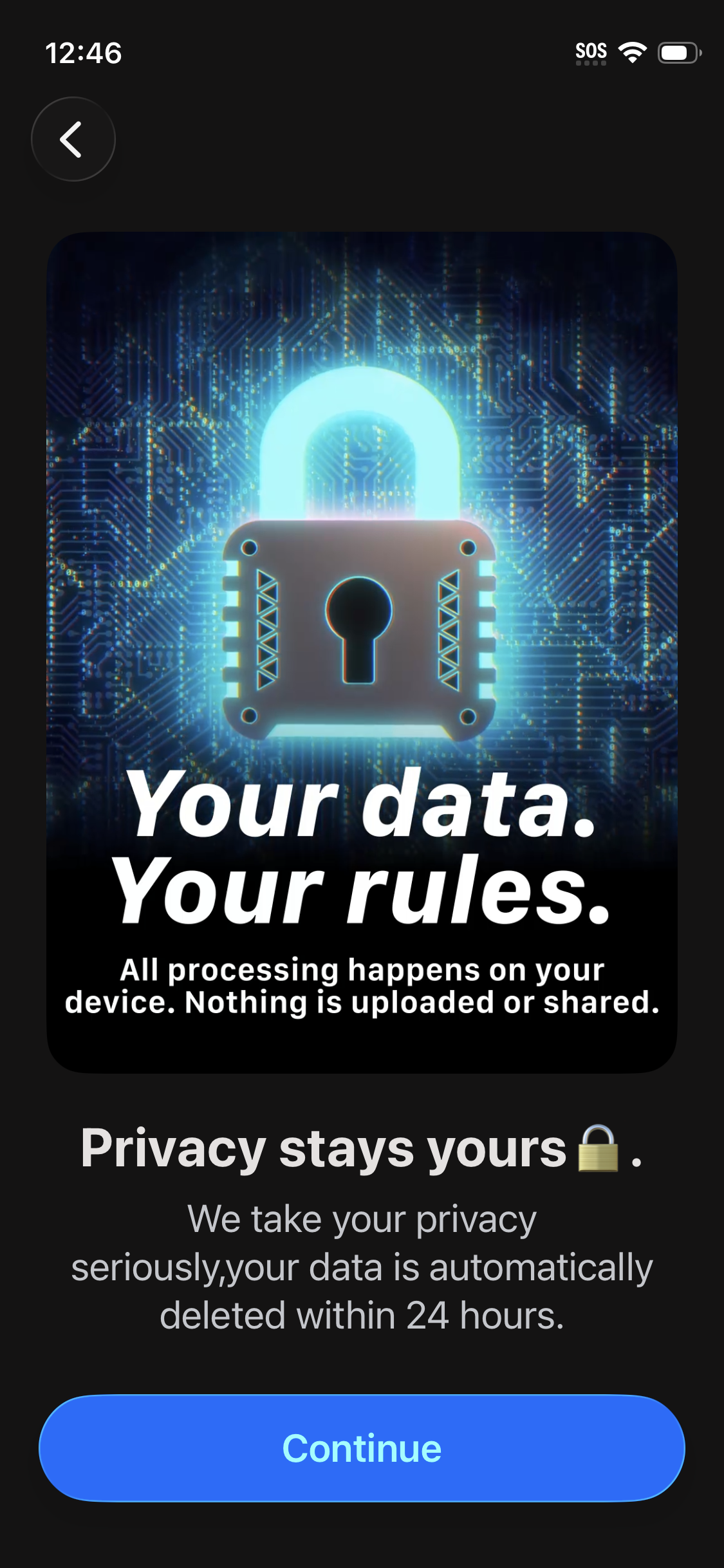}
     \caption{In-app screenshots of \textit{FaceAi: Face Swap Any Video}. The app implies that users can make risque face swaps, and do this privately. These observations motivated our investigation of app descriptions and privacy policies. } 
     \label{fig:screenshot-description}
\end{figure}

\subsection{App Descriptions (RQ2)}
\label{sec:app-descriptions}

In this section, we investigate whether the apps' descriptions on their store listing pages suggest that the app can be used to create SNCII. We hypothesized that apps may tacitly suggest that face swaps can be used for SNCII: anecdotally, while testing we observed a few examples of in-app pages that make suggestive references about motivations for using SNCII, such as the app in \figref{fig:screenshot-description}, which had an in-app tutorial screen that said ``put your face wherever your imagination dares. we won't judge winking emoji'', and that ``all processing happens on our device''. 

To systematically detect whether such suggestions are prevalent, we analyzed the app store descriptions of all apps in our dataset. 
We find that none of the app descriptions in the app store listings include any mention of nudification or SNCII creation. However, our safety testing has shown that the majority can be used for creating nude images of particular individuals. 

Similar to \cite{brigham2026examining}, this confirms the dual-use nature of face swapping apps --- apps are advertised on the app stores for presumably benign purposes such as entertainment and creative use but are capable of being used for harm. Beyond face swapping functionality, we also find that the majority of apps bundle complementary features that enhance realism (e.g., AI filters), extend the output to new modalities (e.g., AI animation, voice cloning), or target a specific real person's identity (e.g., celebrities). These bundled capabilities make it possible for UI-bound users to create SNCII with multiple modalities from a single image, such as animating and voicing a face-swapped nude image.

\subsubsection{Face swap apps' feature sets}
The majority of face swap apps (n=106, 68.3\%) are usually bundled with one or more other features such as image editing, AI filters and effects (e.g., makeup, beauty filters, haircuts), and AI voice (e.g., voice cloning). Table~\ref{tab:faceswap_features} summarizes the types of additional features. 

\paragraph{Features for multimodal deepfakes.}
We find that the most common features bundled with face swaps are AI filters and image effects (n=70) which enhance the realism of face swap outputs by adjusting skin tone, lighting, and facial expressions. For example, Face Swap - AI Photo Editor, 
an Android app advertises: \textit{``Beyond face swapping, the app offers a suite of AI-powered editing tools to elevate your photo experience. Sharpen blurry images, create stunning AI-generated art, or use the Face Changer to tweak your appearance with precision.''} 
Additional screenshots of app descriptions are available in the Appendix, see \figref{fig:description-sample}). 
We find that 72.9\% of apps advertising AI filter \& image effects in their descriptions are unsafe.

Beyond still image manipulation, 27 apps offer AI animation, enabling the generation of animated face swap content (see \figref{fig:description-sample}): \textit{``Animate your pictures — bring them to life with AI‑powered animations.''} Another six apps include AI voice cloning, allowing users to pair a swapped face with a synthesized voice, 83.3\% of which are unsafe apps. One other app additionally advertises an AI kiss feature: \textit{``AI Kiss video simply creates a live kissing video animation of anyone's photos.''} The inclusion of such features situates these apps within the broader ecosystem of AI companionship applications, which have been shown to enable the simulation of intimate interactions with real people's likenesses without their consent \cite{brigham2026examining}. Additionally, prior work has shown that due to cultural differences such content is perceived to be problematic \cite{amna2024ncida}.

Taken together, these feature combinations suggest that many apps marketed as \textit{``face swappers''} offer additional capabilities that extend face swapping to create more realistic identity replacements (i.e., deepfakes).

\paragraph{Features for face morphing.}
Another class of features we find bundled with 47 face swap apps is face morphing. Instead of swapping one real identity (face) into another, these features support the creation of new facial identities based on one or more input images. Some features included baby face generators, gender swaps, and swapping individual facial features (eyes, nose, etc.). Our safety evaluation has shown that 72.3\% of apps advertising face morphing features are unsafe.

\begin{table}[t]
\centering
\begin{tabular}{lrr rr}
\toprule
\textbf{Feature} & \textit{n} & \textit{\%} & \multicolumn{2}{c}{\textbf{Unsafe}} \\
\cmidrule(lr){4-5}
 & & & \textit{n} & \textit{\%} \\
\midrule
AI filters \& image effects          & 70 & 66.0 & 51 & 72.9 \\
Face morphing                        & 47 & 30.3 & 34 & 72.3 \\
AI animation                         & 27 & 25.5 & 17 & 63.0 \\
Stylized photos                      & 27 & 25.5 & 15 & 55.6 \\
Image editing                        & 24 & 22.6 & 16 & 66.7 \\
Other GenAI                          & 17 & 16.0 &  7 & 41.2 \\
AI clothes/outfit change             & 12 & 11.3 &  7 & 58.3 \\
AI voice                             &  6 &  5.7 &  5 & 83.3 \\
AI kiss/boyfriend/girlfriend         &  1 &  0.9 &  1 & 100.0 \\
\bottomrule
\end{tabular}
\caption{Additional features co-occurring with face swap (n=106 apps mentioning face swap alongside other features). A single app can have multiple features. Of apps that advertise each feature, we report the percentage that produced unsafe outputs during safety testing.}
\label{tab:faceswap_features}
\end{table}

\paragraph{Face swap with no other features.} We find that 47 apps included a description of face swap functionality in isolation. The term \textit{face swap} on app descriptions encompasses a broad range of capabilities, including custom face swap but also template-based face swap, video and live swaps, and group face swaps. For example, Vidqu, an iOS app is advertised as a \textit{``Video Face Swap app that helps you merge any faces in photos and videos!''}. Another popular app Reface, emphasizes swapping into existing templates: \textit{ ``Tap and Swap into anyone: put your face into trending videos, movie stills, or character art — just pick a picture from our collection and go.''} However, both apps offer custom face swaps resulting in unsafe outputs. 


\paragraph{No mention of face swaps.} Two apps do not mention that the app can be used to complete face swaps in their description even though we were able to perform a face swap during testing. Both apps instead advertise AI image and video creation, for example: \textit{``... goddess pose, pure ethnic clothing, charming male god, fashion street shot''} which promotes the use of AI-generated templates for \textit{``one click production of popular videos''}.


\begin{table*}[t]
\centering
\resizebox{0.8\textwidth}{!}{%
\begin{tabular}{lrr rr rr rr rr}
\toprule
\multirow{2}{*}{\textbf{Category}} & \multicolumn{2}{c}{\textbf{Total}} & \multicolumn{2}{c}{\shortstack{\textbf{Also mention}\\\textbf{sharing}}} & \multicolumn{2}{c}{\textbf{Unsafe}} & \multicolumn{2}{c}{\shortstack{\textbf{Partial}\\\textbf{Safety}}} & \multicolumn{2}{c}{\textbf{Safe}} \\
\cmidrule(lr){2-3} \cmidrule(lr){4-5} \cmidrule(lr){6-7} \cmidrule(lr){8-9} \cmidrule(lr){10-11}
 & \textit{n} & \textit{\%} & \textit{n} & \textit{\%} & \textit{n} & \textit{\%} & \textit{n} & \textit{\%} & \textit{n} & \textit{\%} \\
\midrule
No clear motivation                & 23 & 14.8 & \multicolumn{2}{c}{—} & 13 & 11.9 &  2 & 12.5 &  8 & 26.7 \\
Entertainment                      & 38 & 24.5 & 30 & 78.9 & 27 & 24.8 &  5 & 31.2 &  6 & 20.0 \\
Identity-targeting: Celebrity      & 63 & 40.6 & 52 & 82.5 & 50 & 45.9 &  7 & 43.8 &  6 & 20.0 \\
Identity-targeting: Friends/family & 45 & 29.0 & 38 & 84.4 & 33 & 30.3 &  5 & 31.2 &  7 & 23.3 \\
\midrule
\textbf{Total}                     & \textbf{155} & & \multicolumn{2}{c}{} & \textbf{109} & & \textbf{16} & & \textbf{30} & \\
\bottomrule
\end{tabular}%
}

\caption{Advertised motivation categories along with the set of apps in each category that support content sharing/distribution, and the category's safety outcome. Besides the first category, remaining categories are not mutually exclusive; an app advertising both celebrity and friends/family targeting is counted in both rows. As a result, counts and percentages sum to more than 155 and 100\% respectively.}
\label{tab:motivation_safety}

\end{table*}


\subsubsection{Advertised uses for face swaps.} 
Beyond the features offered, app store descriptions reveal the use cases that developers promote to prospective users. We identify four intended use categories across the 155 in-scope apps: apps with no clear use case, apps for entertainment, and apps that explicitly promote swapping faces with  specific individuals: either celebrities, or friends/family (which we refer to as ``identity-targeting'').

\paragraph{Entertainment \& creative use.}
Nearly a quarter of apps (n=38) are marketed towards entertainment and creative use, framing face swap as a tool for engaging with pop culture or placing oneself in fictional or real-life scenarios. For example, ChatUp an Android app mentions: \textit{``Perfect for TikTok, cosplay edits, or just exploring alternate realities with a versatile face swap generator.''} The majority of these apps (n=30, 78.9\%) also simultaneously promote sharing the resulting content, suggesting that even entertainment focused apps are oriented towards content distribution rather than private creative use. We find that 24.8\% of unsafe apps promote their use for entertainment. 

\paragraph{Identity-targeting use.}
We also find that a large share of apps explicitly promote identity-targeting use cases (\tabref{tab:motivation_safety}); where an app promotes face swapping into a known identity (e.g., celebrity, friends, family). For example: \textit{``Swap faces with friends, family, or even your favorite celebrities. Create memes and more, and watch as your photos come to life in ways you never imagined.''}
Sixty-three apps (40.6\%) market face swap as a tool for swapping with or into celebrities, while 45 apps (29.0\%) encourage swapping with friends, family, or people the user knows. In both categories, sharing is prominently featured where 82.5\% of celebrity-targeting apps and 84.4\% of friend/family-targeting apps also promote distributing the resulting content.

We also find that a significant number of unsafe apps promote identity-targeting use cases; 45.9\% and 30.3\% of unsafe apps include celebrity and friends/family targeting respectively with some apps appearing in both categories (\tabref{tab:motivation_safety}). Finally, 23 apps (14.8\%) include no explicit motivational framing in their descriptions, focusing solely on technical capabilities without indicating intended use cases. We find that 26.7\% of safe apps have no clear motivation. 

We did not find that the motivations listed in the descriptions were associated with whether the app was safe or not. We conducted a modified Pearson Chi-squared test to account for multiple response categorical variables (each app could be coded with multiple motivations)~\cite{Thomas01032004}, which did not detect a significant association between motivation and app safety ($\chi^2(5.86, N=155)=8.82, n.s.$).

\subsection{Terms of Service (RQ3)}
\label{sec:results_tos}
In this section, we analyze the terms of service documents of face swap apps, coding whether they contain provisions intended to prevent users from using their apps to create SNCII. Acceptable use policies are often the foundation for regulating user activity on a service~\cite{DOHERTY2011201}, so we hypothesized that terms of service that prevent usage related to SNCII creation may be correlated with the safety of the app. Somewhat surprisingly, we find a very low percentage of apps have provisions relating to SNCII, and no correlation between terms of service and whether apps actually have technical safety measures. 
Table~\ref{tab:tos} displays the counts for provisions we analyzed (each of which we will describe below), and how they break down across the safety testing result from Section~\ref{sec:results_swap}.


\subsubsection{Apps with terms of service}
Of the 155 face swap apps we evaluated, we found only 106 (68\%) had a terms of service document linked in the description of the app, or available on the developer's website. A large percentage of apps had no terms of service, which are not required by the App Store or Play Store. This did not correlate with the safety of the app: a two-sample Pearson chi-squared test indicated no significant association between whether an app had a terms of service and whether it had a safety filter ($\chi^2(2,N=155)=1.23, n.s.$).

\begin{table*}[t]

\resizebox{1\textwidth}{!}{%
\begin{tabular}{lccccccccc}
\toprule
\textbf{Safety} & \textbf{Total} & \makecell{\textbf{Has} \\ \textbf{ToS}} & \makecell{\textbf{AI-generated} \\ \textbf{violations}} & \makecell{\textbf{Require} \\ \textbf{consent}} & \makecell{\textbf{Obscene} \\ \textbf{sexually} \\ \textbf{explicit}} & \makecell{\textbf{Abuse} \\ \textbf{harassment}} & \makecell{\textbf{Defamation} \\ \textbf{misinformation}} & \makecell{\textbf{Illegal} \\ \textbf{harmful}} & \makecell{\textbf{IP} \\ \textbf{violations}}\\
\midrule
Unsafe & 109 & 77 (70.6\%) & 13 (11.9\%) & 27 (24.8\%) & 36 (33.0\%) & 46 (42.2\%) & 53 (48.6\%) & 58 (53.2\%) & 60 (55.0\%) \\
Partial safety & 16 & 11 (68.8\%) & 1 (6.2\%) & 4 (25.0\%) & 5 (31.2\%) & 8 (50.0\%) & 9 (56.2\%) & 8 (50.0\%) & 9 (56.2\%) \\
Safe & 30 & 18 (60.0\%) & 3 (10.0\%) & 7 (23.3\%) & 11 (36.7\%) & 12 (40.0\%) & 16 (53.3\%) & 16 (53.3\%) & 17 (56.7\%) \\
\midrule
\textbf{Total} & \textbf{155} & \textbf{106} & \textbf{17} & \textbf{38} & \textbf{52} & \textbf{66} & \textbf{78} & \textbf{82} & \textbf{86} \\
\bottomrule
\end{tabular}
}
\caption{A summary of terms of service documents provisions that pertain to SNCII, comparing safe and unsafe apps. We did not find a significant association between the presence of such provisions and app safety}
\label{tab:tos}
\end{table*}

\subsubsection{Apps specifically mentioning prohibited uses of AI functionality}
We coded whether terms of service documents had provisions that specifically mentioned prohibiting usage of AI functionality to produce harmful content. For example, the FaceSwapper app references deepfake-specific harms: \textit{When you use our related functions involving face editing or synthesis, you should especially comply with: 1) not to synthesize illegal content; 2) not use composite content for illegal or infringing purposes; 3) not lie about composite content as natural content.} 

Overall we found that only 17 apps (16\% of apps with ToS) prohibited uses around AI functionality. A two-sample Pearson chi-squared test indicated no significant association between having specific provisions around the app's AI functionality and whether it had a safety filter ($\chi^2(2,N=155)=0.44037, n.s.$).

\subsubsection{Requiring consent from individuals depicted} 
We coded whether terms of service documents require users to obtain consent from people depicted in images used for face swaps. We coded this separately from provisions forbidding other forms of likeness misuse, such as copyright infringement or publicity rights, as concepts are more related to intellectual property law rather than individual privacy. For example, the Face Swap Mash Up app states \textit{You May:
Use photos of friends and family with their permission...
You May NOT:
Use photos without permission: Do not use photos of people without their consent.}

38 apps (36\% of apps with terms of service) included a provision requiring consent from the people to use their image. 
This provision was not associated with the whether the app had safety filters in place, a Pearson Chi-squared test indicated no significant association between provisions requiring consent from people depicted in images used for face swaps and whether the app had safety filters ($\chi^2(2,N=155)=0.094, n.s.$).

\subsubsection{General provisions prohibiting harmful uses} 
Lastly, we analyze whether terms of service included more generic prohibited use provisions, that are not specific to AI or face swap functionality, but may be related to SNCII. Our codes roughly map to provisions that are common in other software acceptable use policies: harmful, illegal, and copyrighted content~\cite{pater2016harassmentpolicy,weidman2019acceptable,DOHERTY2011201,schaffner2024communityguidelines}. We describe the provisions we found, and how they link to SNCII.

\paragraph{Intellectual property} 86 apps (55\%) prohibit using the app to violate intellectual property rights such as copyright infringement, or publicity rights. This could apply to explicit face swaps in certain cases, like if a copyrighted image of a celebrity is used, or if a sex worker's image is used to create deepfakes that are later sold, which would be an unauthorized commercial use of their identity. 

\paragraph{Illegal conduct} 82 apps (53\%) prohibit using the act for any illegal conduct. If a user distributed explicit face swaps, it would be considered a criminal violation in the U.S., under the TAKE IT DOWN Act.

\paragraph{Defamation, misinformation, or other falsities} 78 apps (50.3\%) prohibit creating content that is defamatory, spreads misinformation or other false claims. Explicit face swaps could be construed defamatory, if the intent of the perpetrator is to make others falsely believe the person depicted was actually involved in explicit acts. 

\paragraph{Harassment or abuse} 66 apps (42\%) prohibited using the app to harass or abuse another person. SNCII is often used by perpetrators to harass the victims in the real world, like in cases of bullying at schools, or posting the images to social media.

\paragraph{Obscene, sexually explicit, pornographic content} 52 apps (33\%) prohibited usage of the app to create explicit content. Most SNCII is explicit content by definition, but some cases of intimate imagery, such as face swaps meant to simulate kissing or relationships, may not contain nudity.

Similar to the other provisions we coded, we found that this collection of provisions was not correlated with whether the app had safety filters. Because all of these provisions often co-occur as legal boilerplate, we conducted a two-sample Pearson Chi squared test to detect associations between whether apps had safety filter, and whether any of the five above provisions was present in the terms of service. We did not find a significant association between these variables ($\chi^2(2, N=155)=3.02, n.s$).

\subsubsection{Takeaways}
We found that face swap apps often do not have terms of service documents, and even among the ones that do, around half or less contain provisions that prohibit their users from creating SNCII. Morever, the apps that do have such provisions are not more likely to prevent SNCII creation at a technical level with safety filters, than apps without these provisions. This suggests that as a whole, face swap app developers are not deeply concerned about preventing SNCII as a matter of policy. Furthermore, it shows that the content of terms of service documents are not predictive of the safety of the app.

\subsection{Privacy Policies and Practices (RQ4)}
\label{sec:results_privacy}

We analyzed face swap apps' privacy policy documents to examine whether apps' privacy practices enable users to generate SNCII privately. Based on anecdotal observations of in-app descriptions (see \figref{fig:screenshot-description}), we initially hypothesized that face swap processing would predominantly occur locally on-device, as this would reduce developer liability.

We labeled provisions of privacy policies pertaining to how the app handles the images submitted to the app for face swap, and the generated outputs, including whether such data was mentioned at all, where the data was processed, how long it was retained, and whether the developers had the right to use or share users' images. Table~\ref{tab:privacy_policy} summarizes the counts of the labels.
Additionally, we analyze whether provisions are more likely to appear in safe or unsafe apps, using Fisher's exact tests for each provision, with p-values corrected using the Holm-Bonferroni procedure.
Of the 155 apps we tested, only 133 had privacy policies (App=73, Play=60). Additionally, of the 133 apps with privacy policies, eight documents had formatting errors (e.g., malformed documents, non-English ones, error pages).

\subsubsection{Biometric data disclosure.} We examined whether in-scope apps explicitly acknowledged the collection and processing of biometric or face image data in their privacy policies. Some apps (n=27) in our dataset relied on generic, templated privacy policies that makes no reference to biometric data or face images despite it being the core functionality of the app. 

Of the 133 in-scope apps, 98 apps explicitly mentioned biometric or face data in their privacy policy. This includes facial images or biometric data such as facial geometry. We find that both the majority of unsafe (62.4\%) and  safe apps (70\%) discuss the collection and processing of facial data. Fisher’s exact test indicated no statistically significant association between the app's safety result and whether the privacy policy mentioned biometrics ($p_{adj}=1.00$).”

\subsubsection{Data processing \& storage.} We wanted to learn where apps process face image data, distinguishing between cloud-based and on-device (local) processing. We find that the majority of apps (n=81) relied on cloud-based processing. Only 17 apps processed face data locally on the device. There was no difference in processing location between safe and unsafe tests; Fisher's exact test did not find a statistically significant association ($p_{adj}=1.00$).

The overwhelming reliance on cloud processing could reflect that these apps are following standard practices with app infrastructure regardless of the features supported by the app. It could also be negligence on the part of app developers.

\subsubsection{Data retention.} We also explored how apps disclosed their data retention practices for face image data. Among unsafe apps, retention policies were distributed across temporary retention (n=27 apps), immediate deletion (n=25 apps), and a substantial proportion that did not mention retention at all (n=34). 
We find that 40\% of safe apps were more likely to report temporary retention. On the other hand, partially safe apps showed the highest proportion of indefinite retention (n=2 apps), though the small sample size limits interpretation. Fisher's exact test did not find a significant association between safety and the retention policies ($p_{adj}=0.29$).

\subsubsection{Image rights.} We also examined whether apps claimed the right to use users' images, or share users' images with third parties. Among all in-scope apps, 80 (54.1\% of unsafe, 50.0\% of partially safe, and 43.3\% of safe apps) did not claim any rights over face images. Only 8 apps explicitly claimed such rights, 62.5\% of which are safe apps. The remaining 45 apps did not mention image rights at all. Fisher's exact test did not find a significant association between policies around image sharing and safety outcome ($p_{adj}=0.07$).

Overall, the low prevalence of image usage claims across all groups indicates that most apps either disclaim such rights or simply omit the issue entirely, leaving users without clear information about how their face image data may be used.

\subsubsection{Takeaways}
We found that the majority of apps across all safety outcomes relied on cloud-based processing, suggesting that evasion is not a primary design consideration for app developers. More broadly, privacy policy content showed no consistent relationship with app safety outcomes. One potential explanation is that the computer vision models used by most face swap apps cannot easily be run on-device. This provides a potential avenue to hold developers responsible for SNCII generation, as they provide an internet service to users.

\begin{table*}[t]
\centering
\resizebox{0.65\textwidth}{!}{%
\begin{tabular}{lllclr}
\toprule
\textbf{Provision} & \textbf{} & \textbf{Unsafe} & \textbf{Partial} & \textbf{Safe} & \textbf{Total} \\
\midrule
 & Total in-scope & 109 & 16 & 30 & 155 \\
 & Has Privacy Policy & 95 & 12 & 26 & 133 \\
 & Valid Privacy Policy & 90 & 11 & 24 & 125 \\
\midrule
\multirow{2}{*}{Biometric Data}
 & Yes & 68 (62.4\%) & 9 (56.2\%) & 21 (70.0\%) & 98 \\
 & No  & 22 (20.2\%) & 2 (12.5\%) &  3 (10.0\%) & 27 \\
\midrule
\multirow{3}{*}{Processing Location}
 & Cloud                  & 55 (50.5\%) & 8 (50.0\%) & 18 (60.0\%) & 81 \\
 & Local                  & 13 (11.9\%) & 1 (6.2\%)  &  3 (10.0\%) & 17 \\
 & Biometric Not Disclosed & 22 (20.2\%) & 2 (12.5\%) &  3 (10.0\%) & 27 \\
\midrule
\multirow{5}{*}{Retention Duration}
 & Immediately Deleted    & 25 (22.9\%) & 1 (6.2\%)  &  4 (13.3\%) & 30 \\
 & Retained Temporarily   & 27 (24.8\%) & 5 (31.2\%) & 12 (40.0\%) & 44 \\
 & Retained Indefinitely  &  7 (6.4\%)  & 2 (12.5\%) &  0 (0.0\%)  &  9 \\
 & Not Mentioned          &  2 (1.8\%)  & 0 (0.0\%)  &  2 (6.7\%)  &  4 \\
 & N/A                    &  7 (6.4\%)  & 1 (6.2\%)  &  3 (10.0\%) & 11 \\
 & Biometric Not Disclosed & 22 (20.2\%) & 2 (12.5\%) &  3 (10.0\%) & 27 \\
\midrule
\multirow{4}{*}{Image Use}
 & No Rights Claimed      & 59 (54.1\%) & 8 (50.0\%) & 13 (43.3\%) & 80 \\
 & Rights Claimed         &  3 (2.8\%)  & 0 (0.0\%)  &  5 (16.7\%) &  8 \\
 & N/A                    &  6 (5.5\%)  & 1 (6.2\%)  &  3 (10.0\%) & 10 \\
 & Biometric Not Disclosed & 22 (20.2\%) & 2 (12.5\%) &  3 (10.0\%) & 27 \\
\bottomrule
\end{tabular}
}
\caption{A summary of privacy policy provisions relating to use and processing of facial image data, comparing safe and unsafe apps We did not find significant associations between privacy policies and app safety.}
\label{tab:privacy_policy}
\end{table*}

\section{Discussion}

Our results show that ``dual-use'' face swap apps that do not advertise nudification capabilities are common, and the majority of such apps make no effort to prevent users from creating SNCII. This category of apps may represent a significant source of SNCII created by UI-bound users. 
In this section, we discuss possible approaches to governance for unsafe face swap apps and AI image generation models, beginning with an analysis of the legal context, and ending with recommendations for lawmakers, platforms, and developers.



\subsection{Addressing Unsafe AI Image Editing Tools in (U.S.) Law is Challenging}
\label{sec:discussion_law}
Currently, there is political momentum towards regulating nudification apps through regulation. 
The EU AI Act, still being drafted at the time of this publication, will provisionally prohibit both nudifier apps and dual use AI apps that do not contain safety measures that prohibit creation of SNCII~\cite{europarl2026aiact}, such as the ones we studied. 

However, in the United States, the path to regulating nudification capabilities in dual use apps is more difficult. 
While the TAKE IT DOWN Act 
introduced criminal penalties for individuals and platforms that \textit{distribute} SNCII, statutory and First Amendment issues make it difficult to regulate the AI tools used to \textit{generate} SNCII in the first place and hold developers and platforms accountable. We describe some of these legal challenges to preventing regulation of nudification apps.

\paragraph{App platforms are largely shielded from liability}
Mobile app platforms like Apple and Google by and large are not responsible when users create SNCII using apps on their platform.
Section 230 of the Communications Decency Act protects platforms from liability from illegal communications on their platforms, so long as they make a good faith effort to remove illegal content.

The TAKE IT DOWN Act creates a limited carve-out from Section 230 protections, but this likely does not apply to software like face swap apps.
Online platforms are required to remove SNCII distributed on their platforms within 48 hours after notice, or they face criminal liability.
However, the TAKE IT DOWN Act only covers platforms that provide forums where SNCII may be shared, such as social media platforms.
Google and Apple's app stores may host apps that are capable of generating SNCII, but do not host SNCII directly, meaning that this requirement does not apply. 

The face swap apps we audited in our analysis did not also host user-generated content -- they allow generation of face swaps but do not share with others -- which means they are likely also not considered a covered platform under TAKE IT DOWN, and have no additional responsibility under the law.

\paragraph{The First Amendment may constrain lawmakers' ability to regulate unsafe AI image generation tools} Not all AI-generated or manipulated explicit imagery is illegal. Images created with consent of the people depicted remains legal, and explicit imagery is generally considered protected obscene expression under \textit{Miller v. California}. Statutory bans on generating nude images with AI, bans on software that can produce it, or even targeted regulations requiring safety filters on certain algorithms, could face constitutional challenges under the First Amendment. 
Nevertheless, Minnesota passed a the first state-level law banning nudifier apps in April 2026~\cite{mithani2026minnesota}, which may present a test case of whether such bans will survive in court.

\paragraph{App developers outside of the U.S.}
We found that many developers of face swaps apps are based outside of the U.S. (all but 3 of 70 Android apps). Even if laws prohibiting nudification apps enacted, and are specifically written to cover developers in other countries (e.g.,~ COPPA~\cite{fair2018geolocation}), it may be difficult to seek damages against foreign developers.

\subsection{Recommendations for mitigating risks from apps capable of creating SNCII}
Given the difficulties in regulating dual use apps in the U.S.,
platforms like Apple and Google are in a better position to mitigate the risks that these apps pose through stricter policies. 
We make recommendations for platforms, lawmakers, and developers for implementing platform policies for regulating nudification capabilities.


\paragraph{\textbf{Recommendation 1 }(platforms): Update platform policies to prohibit hosting apps that have AI image generation capabilities without safeguards against nudity generation.}
Currently, Apple's content policies for apps are inadequate for addressing apps that give users access to AI image manipulation tools that could be used for nudification. 
While Apple's App Review Guidelines prohibit apps that include objectionable content in themselves, but do not cover apps that appear benign but can readily be misused to create objectionable content like face swap apps~\cite{apple2026appreview}. 
On the other hand, Google Play has a policy for AI-generated content, which specifically requires apps that generate content with AI to prevent restricted content from being created~\cite{google2026aigencontent}.



\paragraph{\textbf{Recommendation 2} (platforms): Audit AI image generation apps through red-teaming. }
Our results showed that manual testing was necessary to determine whether face swap apps had safeguards against generating explicit content. There were no correlations between the presence of safety filters and the popularity of the app, the description of the app on the store page, terms of service provisions on prohibited use, and privacy policy provisions. 

Given the lack of signals in the app metadata, this suggests that some resources on the trust and safety operations of app platforms should be used to periodically red-team a sample of apps with AI image generation and manipulation features, to check for the existence of safety features. To do these tests, operators may need to use nude imagery. We suggest using AI-generated imagery for this purpose, as in our ethical analysis, this approach poses fewer risks than using real images (including consensually obtained images). A deeper discussion of the ethics is available in Appendix~\ref{sec:ethics}.

\paragraph{\textbf{Recommendation 3} (Lawmakers): Create safe harbor provisions for red-teaming with AI-generated nude imagery}
A potential roadblock to red-teaming AI image editing apps with AI-generated nude imagery within platforms are concerns about the legality of this method. We suggest that lawmakers pass safe harbor legislation that exempts good faith trust and safety operations from potential legal risks relating to the TAKE IT DOWN Act. This is similar to calls to create safe harbor exemptions for security researchers under the CFAA~\cite{pfefferkorn2022shooting}, and for AI safety researchers evaluating the risk that image models create AI-generated CSAM~\cite{pfefferkorn2026aigrok}.



\paragraph{\textbf{Recommendation 4} (developers): Apps with AI image generation and manipulation features should implement nudity filters and deterrence messaging to discourage SNCII generation}
Apps that use AI-based tools for image generation and manipulation run the risk of abuse via SNCII, whether the developers intend the app to be used for creating SNCII or not. Developers who do not intend for their apps to be used to create explicit imagery should implement nudity detection filters. Developers who do intend for their app to be used to create explicit images should take additional measures to discourage non-consensual image generation and manipulation. For example, using deterrence messages to inform users of the consequences and risks of non-consensual usage of others' images~\cite{rao2025deterrence}.

\paragraph{\textbf{Recommendation 5} (litigators): hold platforms and apps responsible when they do not implement safeguards}
Litigators could play a role in pressuring app developers and platforms to put in place safeguards against SNCII, by holding them accountable when they are clearly negligent.
While Section 230 grants online platforms broad immunity for content generated by users, platforms may lose protection if they "materially contribute" to making illegal content, rather than simply providing a neutral tool that users misuse. There is precedent for such an approach: in \textit{Fair Housing Council v. Roommates.com}, a roommate-finding platform was denied immunity because it required users to disclose preferences for protected characteristics (such as sex, sexual orientation, and family status) which violate fair housing law. By requiring such responses as a platform, it no longer was neutral~\cite{eff2008roommates}. We argue that an AI image-editing app could lose its immunity if it does not implement safeguards against creating SNCII, and if the marketing and design of the app suggests or encourages the creation of SNCII (such as in \figref{fig:screenshot-description}).

\subsection{Limitations}
Lastly, we discuss limitations of our methods and possible implications for our results.
We conducted a keyword search on the app stores using a single keyword \textit{face swap}. This might have only surfaced a subset of face swapping apps, or surfaced more relevant apps in one app store compared to the other. We limited our safety testing to a two-image input custom face swap feature, excluding apps that had video and live face swapping features, which means our results likely undercount the true number of unsafe face swap apps. It is also possible that some apps detect AI-generated images, which could have influenced our safety evaluation outcomes; for instance, by triggering stricter or more lenient safety filters than would apply to real images.

\section{Conclusion}
In this work, we found that 70\% of face swap apps on the iOS App and Google Play stores permit users to generate face swaps using explicit images. Furthermore, none of these apps are advertised as nudification apps. This suggests that face swap apps, and many other forms of AI image generation and editing apps, are effectively ``dual-use'': apps that evade content moderation by platforms because they present as benign, but possess the capability to create harmful content like SNCII. Such apps may represent a large portion of the tools used by UI-bound users to create SNCII today. To combat the easy accessbility of tools for creating SNCII, we suggest that all platforms that host applications that include AI image generation and editing functionality institute more stringent content policies and conduct periodic safety testing.


\newpage
\bibliographystyle{plain}
\bibliography{references}

\appendix

\section{Ethical Considerations}
\label{sec:ethics}

\subsection{Use of Generated Nude Images}

To rigorously test whether face swap apps implement safety measures to prevent non-consensual intimate imagery creation, we decided to use a small set of AI-generated nude images. At a high level, our testing procedure involves: 1) creating a clothed ``target face'' image and a nude ``target body'' image. 2) For each app, we attempt to swap the target face onto the target body.

The usage of nude images (authentic and generated) in research is ethically fraught~\cite{cintaqia2025stop}. In this section we discuss the ethical issues involved, and how we arrived at our testing methodology of using a small set of AI-generated images.

We identified two primary harms involved with using nude images to test face swap apps, that any method would need to address:
\begin{enumerate}
\item \textbf{Harm 1: Consent to Face Swapping} Swapping a face onto a nude body, if done without the consent of the people in both images, is a violation of both peoples' privacy, bodily autonomy, and research ethics on informed consent. Any study using nude images of real people would require consent, at minimum. 
\item \textbf{Harm 2: Harmful usage by third parties.} Even if obtained with consent, the images used for testing may be collected by app developers. The app developers may use the images for purposes that people did not consent to, such as generating their own nonconsensual intimate imagery, or training models. 
\end{enumerate}

\subsubsection{Approaches considered}
Based on these harms, we immediately ruled out using nude or face images obtained from the internet or datasets without consent. We then evaluated two approaches to sourcing nude images that could address these harms. 

\textbf{Approach 1: use images with permissive licenses} 
We would find nude images in the public domain, licensed with Creative Commons, or other permissive licenses that allow ``remixing'' or transformative uses. 

While this approach is potentially legal from a copyright perspective, the people depicted in the images did not explicitly consent for their likeness or bodies to be used in face swaps. Because this usage is substantially more privacy-invasive and obscene, we determined that the choice of license was not an appropriate substitute for consent, and thus, this approach does not address Harm 1. 

\textbf{Approach 2: partner with nude models or sex workers to consensually obtain input images}. We would pay for or request permission to use their images for the study. We would specifically inform them that their face would be used for a face swap, and that third-party app developers would potentially gain access to their images. 

While this approach appears to address Harm 1 by obtaining explicit consent, we believed that there remained substantial risks of harm. 
First, regarding Harm 1, there remains a risk of financial coercion through participant payments. We consulted with an expert in participatory research with sex workers, who raised the issue that sex workers who were in financial need may participate in the study to receive compensation, despite not feeling truly comfortable the risks. 
Second, regarding Harm 2, we would not be able to predict what third party app developers would do with the images provided by sex workers, and could not protect them from negative consequences.


\textbf{Approach 3: use AI-generated images.} We would use an image generation model to create nude target body images, and target face images. 
This mitigates Harms 1 and 2 because the AI-generated images would not depict real, identifiable people, so no specific individuals would be impacted. 

However, this approach does pose its own harms because of the use of image generation models.
Image models capable of generating nude images are most likely trained on explicit images scraped from the web.
The collective people depicted in the images are harmed by the non-consensual usage of their image for model training and generation, particularly if the images themselves were captured and distributed non-consensually. As a downstream user of the model outputs, we would be benefitting from the harms that enabled these image models to be built.

Furthermore, there is a chance that an AI-generated face or body closely resembles a real person, who would be harmed by the usage of their likeness.
To ensure that we did not accidentally using real person's likeness, we planned to use reverse image search services to ensure that there were no recognizable people in the image.

\subsubsection{Decision}
Ultimately, we decided on Approach 3 (AI-generated images), based on the following rationale:
\begin{itemize}
\item We decided that Approach 1 (public domain images) did not meet the standard for informed consent.
\item We weighed the harms of Approach 2 (images from participating sex workers) vs. Approach 3. Approach 2 could result in acute harm to individual participants if their images were misused by app developers, while Approach 3 represents a diffuse, \textit{post hoc} harm to the collective group of people whose images were used for training image generation models. We concluded that Approach 3 posed a lower overall risk.
\item We weighed the harm of using AI-generated nude images with the benefits of the study. We concluded that a minimal, diffuse harm to individuals in the training data would be outweighed by findings that could lead to policy changes that prevent people from experiencing computer vision-enabled image-based sexual abuse. 
\item We could not identify feasible alternative approaches that would rigorously test whether face swap apps have safety mechanisms to prevent the creation of SNCII; such as reverse engineering or gaining access to insider information from developers.
\end{itemize}

We acknowledge that nevertheless, there are ethical issues with using AI-generated imagery. We attempted to mitigate these issues with the following measures:
\begin{enumerate}
\item Generating as few images as necessary to conduct a scientifically valid study (8 total: 2 genders x 2 skin tones x 2 images per swap-pair). 
\item Using reverse image search to minimize the probability that the images we generate depicted real people.
\item Handling the data sensitively as possible to minimize harm: we will delete the generated and face-swapped images after the conclusion of the study. We will not sharing the generated images beyond the study team.
\end{enumerate}

\subsubsection{Ethics Review}
Because we used AI-generated images, we did not collect any of our images or data through interactions or interventions with human subjects. Thus, to the best of our knowledge, this research does not qualify as human subjects research, and we did not submit our study for review by our IRB.

However, to review the ethics of our study,
we solicited feedback from multiple experts in image-based sexual abuse, online trust and safety, and computer ethics. We worked with these experts throughout the study design process to help enumerate possible harms and risks and provide perspective on our decision-making. Lastly, prior to submission, a computer ethics course provided a review of our project and enumerated any additional ethical considerations before publication, which helped us identify follow-up actions based on our findings.

\subsection{Funding Face Swap Apps}
Many of the apps we identified require a subscription to use. By doing so, it provides financial support to developers who may be enabling image-based sexual abuse. However, as external researchers, we found no alternative methods for evaluating safety in apps that require subscriptions for use. We decided that the potential harm of spending a small dollar amount on subscriptions was outweighed by the potential benefits of our findings. We attempted to spend as little possible to complete the testing, taking advantage of free trials if available, and setting a maximum price threshold of \$20. In total, we spent \$1057 on subscriptions to apps. To offset potential impacts of this spending, we plan to donate an equivalent amount to the Cyber Civil Rights Initiative, a non-profit organization which runs a helpline for victims of image based sexual abuse.

\subsection{Responsible Disclosure}
Neither Apple nor Google's platform policies explicitly forbid apps that can be used for the production of AI-generated nude imagery. Nevertheless, given the likelihood that the face swap apps we identified without safety filters could be abused to create SNCII, we have started the disclosure process to report our findings to Apple and Google.

\section{Open Science}

We make the following artifacts available for our paper using this link: \url{https://anonymous.4open.science/r/face-swap-auditing-D516/}
We make available (1) iOS (n=210) and Android (n=250) app metadata and (2) app metadata scraping scripts. We also include (3) app descriptions, terms of service, and privacy policy codebooks, (4) scraping script for both terms of service \& privacy policies, (5) LLM extraction and labeling prompts. 

We include the (6) app safety testing data for both iOS and Android. We also include the (7) open coded app descriptions used for analysis and the (8) LLM labeled terms of service and privacy policies for all apps, (9) the results from majority voting as described in \figref{fig:tosanalysis}, and the (10) validation results showing the sample terms of service and privacy policies used for intercoder agreement.

\section{Generative AI Usage}

We used the FLUX.2 image generation model to generating the set of test images required for safety evaluation (see \secref{sec:test_images}). We also used Gemini 2.5 Flash for LLM labeling as part of our terms of service and privacy policy analysis, which we describe in detail in \secref{sec:tos_methods} and \secref{sec:privacy_methods}. Beyond that, we used Claude Code to assist with writing the Python and R code required to perform the data analysis on the collected datasets. To validate the correctness of this code, we inspected the code line-by-line as it was generated to ensure that the code matched our intentions, and debugged code by hand if necessary. We used Claude to generate bibtex syntax for some of our citations by providing it with a URL to the source we wanted cited. We manually verified that the bibtex output matched the source.
Lastly, we used Claude for minor editorial improvements to existing written text, such as correcting grammar and spelling.

\section{Generation \& Analysis Prompts}
\label{appendix:test_image_prompts}
As discussed in \secref{sec:methods}, we used the prompts in \tabref{tab:prompts-images} to generate the clothed input face images for all four pairs. Additionally, we used the prompts in \tabref{tab:prompts-extraction} to extract relevant clauses in terms of service documents.

\section{Codebooks}
\label{appendix:codebooks}
We include the codebook (\tabref{tab:describe-codebook}) that guided our analysis of the app descriptions in \secref{sec:app-descriptions}. These are a total of 16 codes across two categories; types of features and face swap use cases. We also include the codebook and set of prompts that guided our terms of service analysis \tabref{tab:tos-codebook} and privacy policy analysis \tabref{tab:pp-codebook}.

\begin{table}[b]
\centering

\resizebox{0.53\textwidth}{!}{%
\begin{tabular}{p{2.5cm} p{8cm}}
\hline
 \textbf{Image} & \textbf{Prompt} \\
\hline

man, light skin & Realistic photograph of a man with light skin, brown hair, face clearly visible, mid 20s, directly facing camera, standing in front of a gray studio background, wearing a button up shirt and pants \\
\hline
man, dark skin & Realistic photograph of a man with dark skin, face clearly visible, mid 20s, directly facing camera, standing in front of a gray studio background, wearing a button up shirt and pants \\ 
\hline
woman, dark skin & Realistic photograph of a woman with dark skin, long straight black hair hair, face clearly visible, thin straight eyebrows, upturned eyes, mid 20s, small nose, directly facing camera, standing in front of a gray studio background, no people in background, wearing a blouse and pants \\
\hline
woman, light skin & Realistic photograph of a woman with light skin, wavy auburn hair, pulled back in a ponytail and face clearly visible, thin straight eyebrows, hooded eyes, mid 20s, large nose, directly facing camera, standing in front of a gray studio background, no people in background, wearing a blouse and pants\\

\hline

\end{tabular}
}
\caption{The prompts we used to generate the clothed input test image set}
\label{tab:prompts-images}
\end{table}

\begin{table*}[h]
\centering

\resizebox{0.7\textwidth}{!}{%
\begin{tabular}{p{0.2\linewidth} p{0.7\linewidth}}
\hline
 \textbf{Document section} & \textbf{Prompt} \\
\hline

prohibited content & This is a terms of service document for an app. Prohibited content is defined as any of the following: Illegal/harmful content; Abuse/harassment; Obscene/sexually inappropriate; Defamation/misinformation; Privacy/personal data violations; IP violations; Security/technical threats; Spam; Content violating laws/regulations; AI-generated content violations; Linking to the app from porn sites. Extract the following from this document verbatim: Every clause or sub-section that explicitly lists or mentions prohibited content matching the definition above, prohibited use cases, restricted activities, or what users must not do. Only include clauses that directly relate to the prohibited content definition above — do not include unrelated restrictions such as age requirements, payment obligations, or technical usage limits. Include only the relevant clause or sub-section and its sub-headings exactly as they appear in the document, not the entire parent section.\\

\hline

user consent & Extract verbatim every clause or sub-section that addresses the requirement for users to have the right to use images or videos of other people. This includes two types of clauses: 1. Explicit consent clauses — clauses that explicitly require users to have obtained consent from individuals depicted in images or videos before uploading or sharing them. 2. Legal rights clauses — clauses that require users to own or hold the necessary legal rights, licenses, or permissions to use the content, including clauses referencing copyright ownership, privacy rights, or publicity rights as they relate to depicted individuals. Do not include general content ownership or licensing clauses that only address the user's own original content. Only include clauses where the rights or consent of another depicted person is implicated. Include only the relevant clause or sub-section and its sub-headings exactly as they appear in the document. \\

\hline

\end{tabular}
}
\caption{The prompts we used to extract the relevant clauses from the terms of service documents}
\label{tab:prompts-extraction}
\end{table*}



\begin{table*}[h]
\centering

\resizebox{0.7\textwidth}{!}{%
\begin{tabular}{l|l|l}
\hline
\textbf{Category} & \textbf{Code} & \textbf{Definition} \\
\hline
\multirow{10}{*}{Types of features} & face swap & Transfers facial features from one person to another \\
& other face identity manipulation & Merges, alters the facial identity of a person \\
& image editing & Traditional image editing techniques pre-deep learning \\
& stylized photos & Image to image style transfer (e.g., AI headshots, business photos, profiles)  \\
& AI clothes/outfit change & Try on different clothes and outfits \\
& AI image effects / "filters" & Effects that modify the face but are intended to preserve identity \\
& AI animation & Create an animated video from an image, may also create a generated voice \\
& AI voice & Voice manipulation only \\
& Other GenAI & "Text-to-Image" or "Text-to-Video" creation of images \\
& AI kiss/boyfriend/girlfriend & Gen AI specifically for intimate stuff that can potentially be abused \\
\hline
\multirow{6}{*}{Face swap use cases} & celebrity & Look like a celebrity \\
& friends/family/people you know & Look like friends, family, and people you know \\
& pop culture (fictional) & Fantasy, trending videos, characters, cosplay \\
& sharing & Share your creations on social media \\
& real-life situations & Template setting such as beach, weddings, travel, etc. \\
& content policy & Any mention of content policy\\
\hline

\end{tabular}
}
\caption{Codebook for app descriptions across two categories: types of features and face swap use cases.}
\label{tab:describe-codebook}
\end{table*}

\begin{figure}[h]
     \centering
     \includegraphics[width=0.7\linewidth]{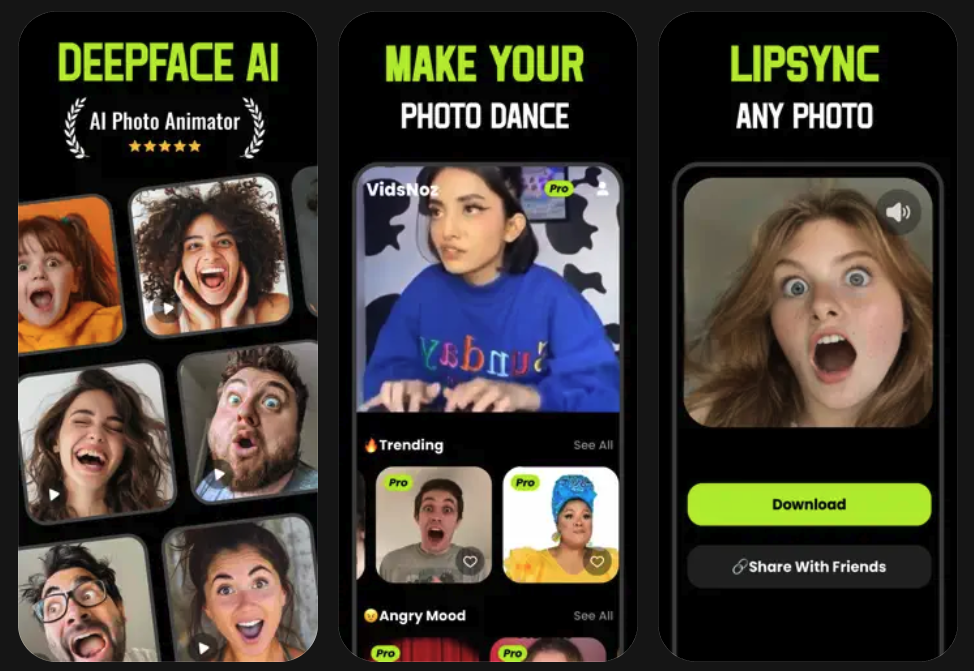}
     \caption{Screenshot from the iOS Deepface AI-Gender Swap Filter app description as seen in the App store listing. The app offers a number of different features: AI animation, AI voice, image editing, and others. This app does not advertise a custom face swap feature, even though it offers one.}
     \label{fig:description-sample}
\end{figure}

\begin{figure}[h]
     \centering
     \includegraphics[width=1.0\linewidth]{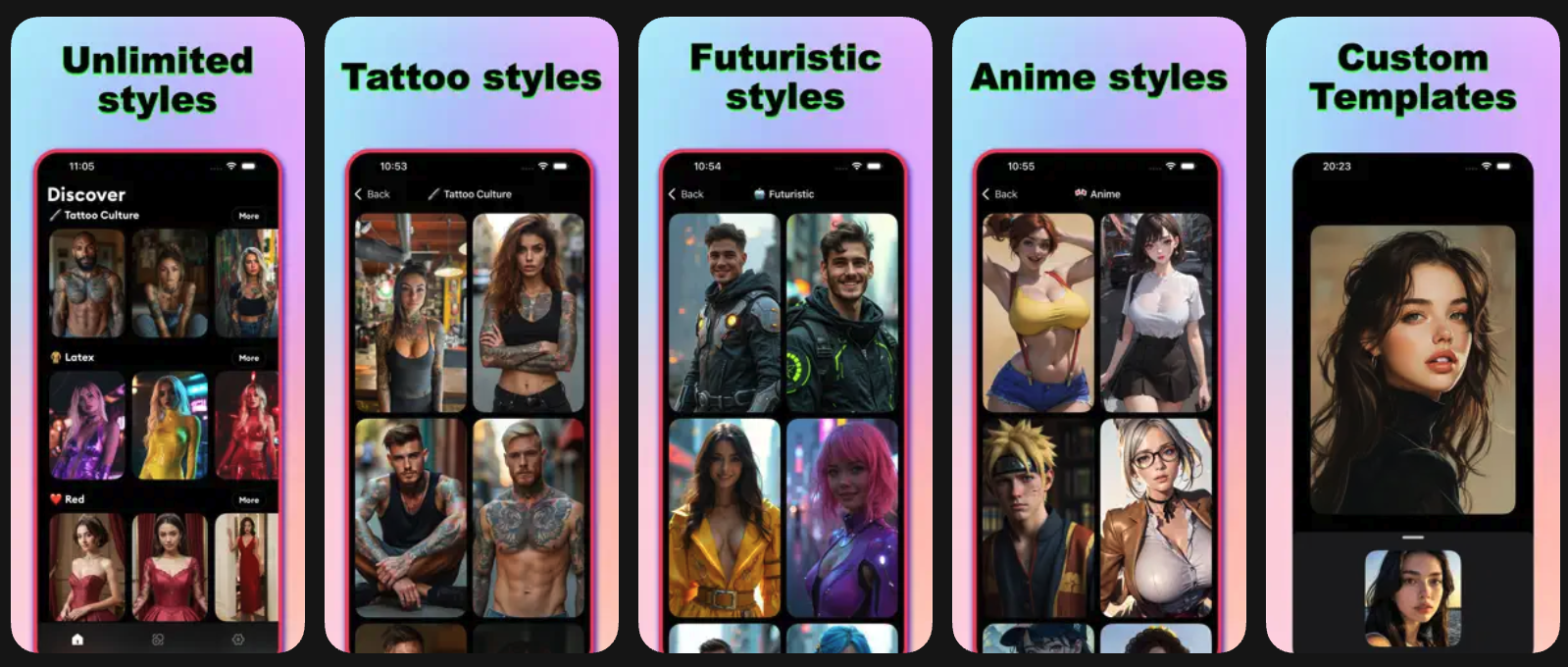}
     \caption{Screenshot showing the different template-based face swaps that are offered by apps. These are templates provided in the AI Face Swap - Replace Face app.}
     \label{fig:description-templates}
\end{figure}

\begin{table*}[h]
\centering

\resizebox{0.6\textwidth}{!}{%
\begin{tabularx}{0.8\textwidth}{p{3.5cm} >{\raggedright\arraybackslash}X}
\hline
 \textbf{Code} & \textbf{Definition} \\
\hline
illegal/harmful & Does this text explicitly prohibit activities that are illegal or prohibited by law? This includes content that violates local and international laws or regulations, criminal activity, weapons development, CSAM, and NSFW content among others? \\
\hline
abuse/harassment & Does this text explicitly prohibit abusive or harassing content directed at other people? \\
\hline
obscene/sexually inappropriate & Does this text explicitly prohibit obscene, sexually explicit, or sexually inappropriate content? \\
\hline
\makecell[l]{defamation/misinformation \\ /impersonation} & Does this text prohibit creating or distributing false statements or representations such as defamation, misinformation, impersonation, etc.?\\
\hline
IP violations & Does this text prohibit intellectual property violations, including copyright infringement, violating a person's publicity rights, personality rights, or using content without the required license or legal rights? \\
\hline
AI-generated content violations & Does this text specifically address AI-generated content violations — for example, prohibiting the use of AI tools and features for the creation and distribution of illegal, harmful, defamatory, abusive, harassing, obscene, sexually inappropriate, false content\\
\hline
link to app from porn sites & Does this text prohibit linking to the app from pornographic or adult websites, or prohibit advertising on such sites? \\
\hline
Requires consent to use images & Does this text require users to have obtained explicit consent from individuals depicted in images or videos before uploading or sharing that content? This is distinct from general IP or licensing clauses — it specifically requires the depicted person's consent. \\

\hline

\end{tabularx}
}
\caption{Codebook for terms of service analysis. The corresponding definition for each code is also the prompt we used for LLM coding and labeling. For each code we asked the LLM to output ``True'' if found and ``False'' otherwise.}
\label{tab:tos-codebook}
\end{table*}

\begin{table*}[h]
\centering

\resizebox{0.8\textwidth}{!}{%
\begin{tabular}{p{0.2\linewidth} p{0.7\linewidth}}
\hline
 \textbf{Code} & \textbf{Definition} \\
\hline
format error & Is this document an invalid privacy policy document? Reasons the document could be invalid include, but are not limited to: the text is malformed, the document is not written in english, or it is an error page. If the document is invalid, output 'true'.\\
\hline
biometric covered & Does this privacy policy specifically address how or whether they handle face image data or biometric data? Answer with ``true'' if handling of face data or biometric data is specifically addressed. Answer with ``false'' if face data or biometric data is not mentioned. \\
\hline
face processing location & Based on this privacy policy, does the app's functionality involve uploading biometric data or face image data to the cloud, or does all processing of such data occur on-device? Answer with ``local'' if all processing of face or biometric data occurs on the user's device, or if the developers claim to not collect any such data. Answer with ``cloud'' if face or biometric data will be uploaded to cloud service providers or on developers' servers. Answer with ``not mentioned'' if the document does not specifically describe where face or biometric data is processed. \\
\hline
retention & Based on this privacy policy, if biometric data or face image data is uploaded developers' servers or in the cloud, is that data retained or eventually deleted? Answer with ``immediately deleted'' if the policy states the data will be deleted immediately after being processed. Answer with ``retained temporarily'' if the policy states the data will be deleted from their servers within a certain duration. Answer with ``retained indefinitely'' if the policy states they will retain the biometric or face data indefinitely. Answer with ``not mentioned'' if the policy does not specifically describe how long biometric or face data will be retained. Answer with ``n/a'' if the data is only processed on-device, or they claim that they do not collect data, and this question does not apply. If multiple types of data are mentioned with different retention policies, answer with the longest retention period mentioned \\
\hline
retention duration & Based on this privacy policy, if biometric data or face image data is retained on developers' servers or in the cloud longer than immediately deleted, how long will the developers retain that data? Respond in exactly this format and nothing else: Answer only with the specific timeframe described in the policy (e.g., ``30 days'', ``1 year'', ``indefinitely''). If there are multiple types of data with different retention periods, answer with a list in the following format: [data type] - [retention period]. For example, 'face data - 30 days; biometric data - 1 year'. Answer with ``not mentioned'' if no timeframe is provided for deletion. Answer with ``n/a'' if the data is only processed on-device, immediately deleted, or they claim that they do not collect data, and this question does not apply. \\
\hline
image use & Based on this privacy policy, if biometric data or face image data is uploaded to developers' servers or in the cloud, does the developer explicitly claim any rights to use face/biometric data for purposes beyond producing the requested output, or to share face/biometric data with third parties? By sharing with third parties, we exclude cloud service providers that provide the infrastructure for running the app. If yes, output ``true''. If they explicitly say the data is not shared or used for other purposes, output ``false''. Answer with ``not mentioned'' if the policy does not specifically address whether face or biometric data will be shared or used for other purposes. Answer with ``n/a'' if the data is only processed on-device, or they claim that they do not collect data, and this question does not apply. \\
\hline
image use types & Based on this privacy policy, if biometric data or face image data is uploaded to developers' servers or in the cloud, and the developer claims rights to use that data for purposes beyond producing the requested output, or says the data may be shared with third parties, list how the data will be used or shared. Respond in exactly this format and nothing else: Answer with a list in the following format: [data type] - [short phrase describing purpose or third party entity]. For example, ``face data - shared with advertisers'' or ``biometric data - used for improving AI models''. Answer with ``n/a'' if they claim that the data is not shared or used for other purposes, or if the data is only processed on-device, or they claim that they do not collect data, and this question does not apply.\\

\hline

\end{tabular}
}
\caption{Codebook for privacy policy analysis. Each code definition corresponds to the LLM prompt used to label the privacy policy documents.}
\label{tab:pp-codebook}
\end{table*}


\end{document}